\documentclass{article}
\usepackage{emulateapj,onecolfloat,graphics}

\begin{document}

\def\vt{\vec{\theta}}
\def\gtorder{\mathrel{\raise.3ex\hbox{$>$}\mkern-14mu
             \lower0.6ex\hbox{$\sim$}}}
\def\ltorder{\mathrel{\raise.3ex\hbox{$<$}\mkern-14mu
             \lower0.6ex\hbox{$\sim$}}}

\twocolumn
[
\title{Clusters of Galaxies in the Local Universe}
\author{C.S. Kochanek${}^1$, Martin White${}^2$,
   J. Huchra${}^1$, L. Macri${}^1$,
   T.H. Jarrett${}^3$, 
   S.E. Schneider${}^4$ \& J. Mader${}^5$ }
\affil{${}^1$ Harvard-Smithsonian Center for Astrophysics, 60 Garden St.,
  Cambridge, MA 02138}
\affil{${}^2$ Departments of Physics and Astronomy, University of California,
  Berkeley, CA 94720}
\affil{${}^3$ Infrared Processing and Analysis Center, 
  MS 100-22, Caltech, Pasadena, CA 91125}
\affil{${}^4$ Department of Astronomy, University of Massachusetts,
  Amherst, MA, 01003}
\affil{${}^5$ McDonald Observatory
  Austin, TX, 78712}
\affil{email: ckochanek@cfa.harvard.edu, mwhite@astron.berkeley.edu,
  jhuchra@cfa.harvard.edu, lmacri@cfa.harvard.edu,
  jarrett@ipac.caltech.edu, 
  schneider@messier.astro.edu, jmader@astro.as.utexas.edu}

\begin{abstract}
\noindent
We use a matched filter algorithm to find and study clusters in both N-body
simulations artificially populated with galaxies and the 2MASS survey.
In addition to numerous checks of the matched filter algorithm, we present
results on the halo multiplicity function and the cluster number function.
For a subset of our identified clusters we have information on X-ray
temperatures and luminosities which we cross-correlate with optical richness
and galaxy velocity dispersions.  With all quantities normalized by the
spherical radius corresponding to a mass overdensity of $\Delta_M=200$ or the
equivalent galaxy number overdensity of $\Delta_N=200\Omega_M^{-1}\simeq 666$,
we find that the number of $L>L_*$ galaxies in a cluster of mass $M_{200}$ is
\begin{displaymath}
 \log N_{*666} = (1.44\pm0.17)+(1.10\pm0.09)\log(M_{200}h/10^{15}M_\odot)
\end{displaymath} 
where the uncertainties are dominated by the scatter created by three choices
for relating the observed quantities to the cluster mass.
The region inside the virial radius has a K-band cluster mass-to-light ratio
of $(M/L)_K=(116\pm46)h$
which is essentially independent of cluster mass.  Integrating over all
clusters more massive than $M_{200}=10^{14}\,h^{-1}M_\odot$, the virialized
regions of clusters contain $\simeq 7\%$ of the local stellar luminosity, quite
comparable to the mass fraction in such objects in currently popular
$\Lambda$CDM models.
\end{abstract}
  ]

\section{Introduction} \label{sec:intro}
 
Clusters of galaxies have become one of our most important cosmological
probes because they are relatively easy to discover yet have physical 
properties and abundances that are very sensitive to our model for the
formation and evolution of structure in the universe.
Of particular interest to us, a cluster sample provides the means to
study the high-mass end of the halo multiplicity function, the average
number of galaxies in a halo of mass $M$, which provides important
insight into the process of galaxy formation.

In this paper we have two objectives.  First, we will demonstrate
the use of a matched filter approach to finding clusters in a redshift
survey using both synthetic catalogs and a large sample of galaxies from
the 2MASS survey (Skrutskie et al.~\cite{Skrutskie97}, Jarrett et al.~\cite{Jarrett00}).
With the synthetic data we can test the algorithm and our ability to
extract the input halo multiplicity function from the output cluster catalog.
Second, we will determine the halo multiplicity function, $N(M)$,
from the 2MASS survey.  This study is the first phase in
a bootstrapping process -- based on a synthetic catalog
known to have problems matching reality in detail we can
calibrate an algorithm which when applied to the real data
can supply the parameters for an improved model of the data.
Then the improved model can be used to improve the algorithm and so on.

In \S\ref{sec:sims} we describe the synthetic and real 2MASS data.
In \S\ref{sec:findcl} we describe our version of the matched filter
algorithm.  In \S\ref{sec:tests} we test the algorithm on the synthetic
catalog, focusing on our ability to determine the halo multiplicity function.
In \S\ref{sec:clusters} we apply the algorithm to the 2MASS sample.
Finally, in \S\ref{sec:conclusions} we discuss the steps which can improve
both the synthetic catalog and the algorithm.

\section{Data: Real and Synthetic } \label{sec:sims}

We searched for clusters using galaxies in the 2MASS survey
(Skrutskie et al.~\cite{Skrutskie97}) and in simulations of the 2MASS survey.
We have chosen the 2MASS survey because of its uniform photometry and large
areal coverage, which makes it ideal for finding the relatively rare rich
clusters in the local neighborhood where a wealth of auxiliary information
is available.

\subsection{2MASS Galaxies}

The 2MASS survey provides J, H, K photometry of galaxies over
the full sky to a limiting magnitude of $K \ltorder 13.75$~mag
(Jarrett et al.~\cite{Jarrett00}).  Based
on the Schlegel, Finkbeiner \& Davis~(\cite{Schlegel98}) Galactic 
extinction model and the available 2MASS catalogs, we selected all 
galaxies with Galactic extinction-corrected magnitudes of
$\hbox{K}\leq 12.25$~mag\footnote{We use the 20~mag/arcsec$^2$
circular isophotal magnitudes throughout the paper.}
and Galactic latitude $|b|>5^\circ$. The final sample contains a total of 
90989 galaxies distributed over 91\% of the sky.
The redshift measurements 
are 89\% complete for $\hbox{K}\leq 11.25$~mag but only
36\% complete between $11.25 < \hbox{K} \leq 12.25$.
The galaxies with redshifts are not a random sample of all of the galaxies --
redshifts are more likely to be of cluster than field galaxies.

With the data corrected for Galactic extinction, the relation between
apparent and absolute magnitudes is
\begin{equation} 
   M_K = K - 5 \log (D_L(z)/r_0) - k(z) = K - {\cal D}(z)
   \label{eqn:absmag}
\end{equation}
where $r_0=10$~pc, $D_L(z)$ is the luminosity distance to redshift $z$ and
$k(z)= -6\log(1+z)$ is the $k$-correction.
The $k$-correction is negative, independent of galaxy type, and valid for
$z \ltorder 0.25$ (based on the Worthey~(\cite{Worthey94}) models).
To simplify our later notation, we introduce
${\cal D}(z) = 5\log (D_L(z)/r_0) + k(z)$ as the effective distance modulus
to the galaxy.  
We use an $\Omega_{\rm mat}=1$ cosmological model for the distances and assume
a Hubble constant of $H_0=100 h$~km/s~Mpc.  For our local sample the
particular cosmological model is unimportant.

Kochanek et al.~(\cite{Kochanek01}) derived the K-band luminosity function
for a subset of the 2MASS sample at comparable magnitudes.
The luminosity function is described by a Schechter function,
\begin{equation}
    \phi(M) = { dn \over d M } 
     = 0.4 \ln 10\ n_* \left( { L \over L_* } \right)^{1+\alpha} \exp(-L/L_*)
\end{equation}
where $M_K= M_{K*} - 2.5\log(L/L_*)$ and the integrated luminosity
function is
\begin{equation}
    \Phi(M) = \int_{-\infty}^M \phi(M) dM = n_* \Gamma[1+\alpha, L/L_*].
\end{equation}
The parameters for the global luminosity function are 
$n_*=(1.16\pm0.10)\times 10^{-2} h^3$/Mpc$^3$, $\alpha=-1.09\pm0.06$
and $M_{K*}=-23.39\pm0.05$ using the 2MASS 20~mag/arcsec$^2$ isophotal
magnitudes.  This luminosity function is consistent with that derived
by Cole et al.~(\cite{Cole01}). 
The total luminosity of a galaxy is approximately 
$1.20\pm0.04$ times the isophotal luminosity.

We should keep in mind that the early-type and late-type galaxies
have different luminosity functions, with the early-type galaxies
being brighter but less common than the late-type galaxies.
The algorithm for finding clusters requires both a field luminosity function
$\phi_f(M)$ and a cluster luminosity function $\phi_c(M)$
which will be different because of the morphology-density
relation (e.g.~Dressler~\cite{Dressler80}).  In our present study we assume
that we can model both using the global luminosity function,
$\phi_f(M)=\phi_c(M)=\phi(M)$.  Once we have catalogs of
2MASS galaxies in different environments we can go back and estimate
the environment-dependent luminosity functions.  Note that
we only need the shape of the luminosity function for the
clusters -- all expressions using $\phi_c(M)$ will appear in 
the ratio $\phi_c(M)/\Phi_c(M_{K*})$, which eliminates
any dependence on $n_*$.

\subsection{Simulations of 2MASS}

We generated a simulated 2MASS catalog based on the approach described in
detail in White \& Kochanek (\cite{WhiKoc}; hereafter WK).
We used as a starting point the $z=0$ output of a $256^3$ particle N-body
simulation of a $\Lambda$CDM `concordance' cosmology whose predicted mass
function agrees well with recent observational estimates
(e.g.~Pierpaoli, Scott \& White~\cite{PSW}).
Halos were identified using the friends-of-friends (FoF) algorithm with a
linking length $b=0.2$ times the mean inter-particle spacing, and galaxies
were `assigned' to halos based on the multiplicity function.
The average number of galaxies assigned to a cluster halo with a FoF mass
$M_{FoF}$ is
\begin{equation}
    N_{FoF}(M) = \left({M_{FoF}\over M_R} \right) + 
      0.7 \hbox{max}\left[1,\left( { M_{FoF}\over M_B } \right)^{0.9}\right]
\label{eqn:nfof}
\end{equation}
galaxies where $M_R=2.5 \times 10^{12} h^{-1} M_\odot$ and
$M_B=4.0 \times 10^{12} h^{-1} M_\odot$ based on the halo occupancy
models of Scoccimarro et al.~(\cite{SSHJ}) which we have modified to better
match the observed number densities of galaxies and the expectation that 2MASS
galaxies will trace the mass more faithfully than a blue selected sample.
We sometimes used the virial mass $M_{200}$ defined as the mass interior to
a radius $r_{200}$ within which the mean density exceeds 200 times the
{\it critical\/} density.
The virial radius, which scales as
$r_{200}=1.6(M_{200}h/10^{15}M_\odot)^{1/3}h^{-1}{\rm Mpc}$,
is slightly smaller than the average radius of the FoF cluster with $b=0.2$,
so that on average $M_{200}=0.66 M_{FoF}$ (median $M_{200}=0.71 M_{FoF}$).
For massive halos the number of galaxies actually assigned to a halo is
Poisson distributed for an expectation value of $N_{FoF}$.
Each halo hosts a `central'  galaxy which inherits the halo center of mass
position and velocity.  Any additional galaxies are randomly assigned the
positions and velocities of the dark matter particles identified with the
cluster by the FoF algorithm.  In this way the satellite galaxies trace the
density and velocity field of the dark matter.

One problem with standard halo multiplicity functions is that they are
one-dimensional functions, $N(M)$, normalized to match a specific sample
(but see Yang, Mo \& van den Bosch~\cite{YanMoBos} for improvements in
this regard).
Even for galaxy mass scales, current models of the halo multiplicity
function lead to too few galaxies to match the luminosity or velocity
functions of galaxies over broad ranges (see Kochanek \cite{Koc};
Scoccimarro et al.~\cite{SSHJ};
Chiu, Gnedin \& Ostriker~\cite{Chiu01}). 
For general theoretical use, we need the luminosity-dependent multiplicity
function, $N(>L|M)$, for the average number of galaxies more luminous than
$L$ in a halo of mass $M$.
For clusters, we might expect a halo multiplicity function to factor as
$N_*(M) \Phi_c(>L)/\Phi_c(>L_*)$ where $\Phi_c(>L)$ is the integrated cluster
luminosity function and $N_*(M)$ is the mass-dependent number of galaxies more
luminous than $L_*$.
In the language of Yang et al.~(\cite{YanMoBos}) this assumption corresponds
to assuming a weak dependence of their $\widetilde{\Phi}_{*}$,
$\widetilde{L}_{*}$ and $\widetilde{\alpha}$ on $M$ in this range.

With our modified multiplicity function, Eq.~(\ref{eqn:nfof}), we can match
the number density of galaxies in the global
Kochanek et al.~(\cite{Kochanek01}) K-band luminosity function or the
K-band galaxy number counts if we view these galaxies as a complete sample
truncated at $L_{\rm min}=0.01 L_*$.
The average number of galaxies assigned to a cluster of mass $M$
at redshift $z$ in a catalog with magnitude limit $K_{\rm lim}$ is
$\langle N(M,z) \rangle = N_{FoF}(M) C(z)$.  The completeness function is
\begin{equation}
   C(z) = { \Gamma[1+\alpha,L(z)/L_*] \over \Gamma[1+\alpha,L_{\rm min}/L_*] }
\end{equation}
where $L(z)=\hbox{max}(L_{\rm min},L_{\rm lim}(z))$ for the limiting
luminosity of $2.5\log (L_{\rm lim}(z)/L_*)=M_{K*}-(K_{\rm lim}-{\cal D}(z))$.
For our parameters, $L_{\rm lim}(z)=L_{\rm min}$ at redshift $cz=1350$~km/s
and $L_{\rm lim}(z)=L_*$ at $cz=14000$~km/s.
The average number of galaxies with $L\geq L_*$ is
\begin{equation}
  N_{*FoF}(M) = { N_{FoF}(M)
   \Gamma[1+\alpha,1] \over \Gamma[1+\alpha,L_{\rm min}/L_*] }
     = 0.042 N_{FoF}(M)
\end{equation}   
for our standard parameters.  On average, clusters with $M h/M_\odot=10^{15}$,
$10^{14}$ and $10^{13}$ are assigned $500$, $53$ and $6$ galaxies of which
$21$, $2$ and $0.2$ are brighter than $L_*$.  Over the range we actually
find clusters, this relation is well fit by a power-law of the form
$\log N_{*666} = A +B \log (M_{200}h/10^{15}M_\odot)$ with $A=1.32$
and $B=0.98$.  We adopt this form to describe the halo occupancy function
and summarize all of our subsequent estimates for it in
Table~\ref{tab:multiplicity}.

The model catalog contained 100706 galaxies, whose clustering properties are
close to those of the observed population.
Unlike WK, where we characterized the observed properties of the galaxy simply
by the accuracy with which its redshift could be determined, we assigned
K-band magnitudes to each galaxy by drawing randomly from the luminosity
function.
This process oversimplifies several aspects of the real survey.
First, we did not include the environment dependence of the luminosity
function.  Second, for lower mass halos where there is essentially one
galaxy per halo, we did not correlate the luminosity with the mass of
the halo.  Third, to mimic the redshift measurements in the 2MASS sample
we assumed that all redshifts were measured for K$\leq 11.25$~mag and that
one-third, randomly selected, were measured between
$11.25<\hbox{K}\leq 12.25$.
This mimics the statistics of the real data, but the redshift 
measurements in the real data are more likely to be of cluster
galaxies than of field galaxies.  
Finally we ignored the $|b|>5^\circ$ latitude cut made for the 2MASS
data and simply analyzed the synthetic catalogs for the whole sky.
Most of these simplifications should make it more difficult to identify
clusters in the synthetic catalog than in the real data.  Indeed, the 
visual impression when comparing slices of the model survey and 
2MASS data is that the ``fingers of god'' are weaker in the model.
We intend to significantly improve our mock galaxy catalogs in the next
iteration, using the relations we derive from the data in later sections
(see \S\ref{sec:conclusions}) combined with a better underlying simulation.

\section{Finding and analyzing the clusters} \label{sec:findcl}

Large cluster samples have been constructed using five general approaches.
In the first method, used to construct the original large catalogs and their
successors (e.g.~Abell~\cite{Abe58}; Shectman~\cite{She85};
Dalton et al.~\cite{APM}; Lumsden et al.~\cite{EDCC};
Ostrander et al.~\cite{ONRG}; Scoddeggio et al.~\cite{Sco99};
Gladders \& Yee~\cite{GlaYee}; White et al.~\cite{WBL}),
clusters were selected as overdensities in the projected density of galaxies
on the sky.  It was quickly realized such surveys suffer from projection 
effects and there is a
large scatter between optical richness and cluster mass
(for recent theoretical studies see
e.g.~van Haarlem, Frenk \& White~\cite{vHaFreWhi}; 
Reblinsky \& Bartelmann~\cite{RebBar}; WK).

Recently, more quantitative versions of these methods based on matched filter
algorithms to find either the discrete cluster galaxies
(Postman et al.~\cite{MF}; Kepner et al.~\cite{Kepner99};
Kim et al.~\cite{Kim01})
or the excess luminosity from unresolved galaxies
(Dalcanton~\cite{Dal96}; Zaritsky et al.~\cite{Zaritsky97};
Gonzalez et al.~\cite{LCDCS})
have been developed to find clusters at intermediate redshifts.
A host of approaches have recently been applied to the data from the SDSS
(see Bahcall et al.~\cite{SDSS} for recent discussion and
 Goto et al.~\cite{Goto01} and Nichol et al.~\cite{C4} for early work).

Second, with the advent of large redshift surveys, cluster catalogs were
constructed by finding three-dimensional overdensities in the galaxy
distribution using the friends-of-friends (FoF) algorithm
(e.g.~Huchra \& Geller \cite{HG82}; Geller \& Huchra \cite{GH83};
Ramella et al.~\cite{Ramella94}; Ramella, Pisani \& Geller \cite{Ramella97};
Christlein~\cite{Christlein00}, Ramella et al.~\cite{Ramella02}).
Studies of the FoF algorithm using N-body simulations
(Nolthenius \& White \cite{Nolthenius87}; Frederic \cite{Frederic95};
 Diaferio et al.~\cite{Diaferio99})
demonstrate the FoF algorithm works well for appropriate choices
of the linking parameters.  The tree or ``dendogram'' methods 
(e.g.~Tully~\cite{Tully87}) are related to FoF methods.

The third approach to building cluster catalogs is to use X-ray surveys
(Gioia et al.~\cite{Gio90};
 Edge et al.~\cite{Edg90};
 Henry \& Arnaud~\cite{HenArn};
 Rosati et al.~\cite{Ros95};
 Jones  et al.~\cite{Jon98};
 Ebeling et al.~\cite{Ebe98};
 Vikhlinin et al.~\cite{Vik98};
 Romer et al.~\cite{Rom00};
 Henry~\cite{Hen00};
 Blanchard et al.~\cite{BSBD};
 Scharf et al.~\cite{ROXS};
 B\"ohringer et al.~\cite{REFLEX};
 Ebeling, Edge \& Henry~\cite{EbeEdgHen};
 Gioia et al.~\cite{Gio01}).
X-ray surveys avoid\footnote{But see Lewis et al.~(\cite{LSEG}).} many of
the problems in selection and characterization which have made optical
catalogs difficult to use, but frequently lack the sensitivity to detect and
characterize groups or clusters at $z\ga 0.5$.
Finally, two rapidly developing approaches are to use surveys based on the
Sunyaev-Zel'dovich effect (Carlstrom et al.~\cite{JC}) or on weak gravitational
lensing (Schneider \& Kneib \cite{SchKne};
Wu et al.~\cite{WCFX};
Hattori et al.~\cite{HKM};
Wittman et al.~\cite{Witetal01};
M\"oller et al.~\cite{MPKB};
but see White, van Waerbeke \& Mackey~\cite{WvWM}).

\subsection{The matched filter algorithm}

We identify groups and clusters in the 2MASS catalog using the modified
optimal filter method we developed in WK, which is itself derived from
the similar algorithms of Kepner et al.~(\cite{Kepner99}) 
and Postman et al.~(\cite{MF}).
These methods identify clusters as 2- or 2.5-dimensional overdensities
in three-dimensional redshift surveys.  This is a departure from the
essentially uniform use of the FoF algorithm to find clusters in redshift
surveys.  While the FoF algorithm clearly works to identify groups and
clusters with very small numbers of galaxies, a matched filter approach
provides several advantages.
Matched filters provide likelihood estimates for the detection, membership
probabilities for individual galaxies, a range of cluster properties and
their  uncertainties and (importantly for us) can naturally incorporate
galaxies both with and without redshift measurements.

In WK we used N-body simulations to explore the selection of clusters in a
series of model surveys for clusters at intermediate redshift.
In particular, we developed strategies for selecting samples above a fixed
mass threshold over a range of redshifts, showing that it was relatively
easy to obtain highly complete samples at the price of significant
contamination from real but somewhat less massive clusters.  Contamination
by detections of non-existent clusters was rare.

The algorithm works as follows.
We model the universe as a uniform background and a distribution
of $k=1 \cdots n_c$ clusters.  Cluster $k$ is described by its 
angular position $\vt_k$, (proper) scale length, $r_{ck}$, a fixed
concentration, $c$, redshift 
$z_k$ and galaxy number $N_k$.  To ease comparisons to theoretical
models we describe the galaxy density profile by an NFW
(Navarro, Frenk \& White~\cite{NFW}) profile.  In three
dimensions, the galaxy distribution normalized by the 
number of galaxies inside an outer radius $ r_{\rm out} = cr_c$ is 
\begin{equation}
   \rho(r) = { 1 \over 4 \pi r_c^3 F(c) } { 1 \over x (1+x)^2 }
\end{equation}
where $x=r/r_c$, $F(x)=\ln(1+x)-x/(1+x)$ and
$4\pi \int_0^{c r_c} r^2 dr \rho \equiv 1$.  The relative
number of galaxies inside radius $r$ is $N(<r)=F(r/r_c)/F(c)$. 
The projected surface density is more complicated, 
\begin{equation}
    \Sigma(R) = { f(R/r_c) \over 2 \pi r_c^2 F(c) }
    \label{eqn:surfden}
\end{equation}
where 
\begin{equation}
     f(x) = { 1 \over x^2-1 } \left[ 1 - { 2 \over |x^2-1|^{1/2} }
         \hbox{tann}^{-1} \left| { x -1 \over x + 1 } \right|^{1/2} \right]
\end{equation}
and $\hbox{tann}^{-1} (x) = \tan^{-1} (x)$ ($\hbox{tanh}^{-1}(x)$) for $x > 1$ 
($x<1$).  The relative number of galaxies projected inside cylindrical
radius $R$ is $N(<R)=g(R/r_c)/F(c)$ where
\begin{equation}
    g(x) = \ln (x/2) + { 2 \over |x^2-1|^{1/2} } 
      \hbox{tann}^{-1} \left| { x -1 \over x + 1 } \right|^{1/2}
    \label{eqn:encmass}
\end{equation}
(Bartelmann~\cite{Bart96}).  The projected quantities are considerably
more complicated than the simple model ($\Sigma\propto (1+x)^{-2}$) we 
used in WK, but the earlier model corresponds to a three-dimensional density
which is too complicated for convenient use.
We fixed the halo  concentration to $c=4$ and the break radius to
$r_c=200 h^{-1}$~kpc, as typical parameters for cluster-mass halos.
We estimate the probability a survey galaxy has its observed properties
as either a field or a cluster member galaxy.  These properties can include
the flux, redshift, angular position and color, with a basic distinction
being made between galaxies with redshifts and those without. 

\subsection{Probabilities for Field Galaxies} \label{sec:fieldprob}

The probability of finding a field galaxy of magnitude $K_i$ and redshift
$z_i$ is
\begin{equation}
  P_f(i) = 0.4 \ln 10\ D_C^2(z_i) { d D_C\over d z} \phi_f(K_i-{\cal D}(z_i))
\end{equation}
where $D_C$ is the comoving distance.
The probability of finding a field galaxy of magnitude K$_i$ is 
\begin{equation}
    P_f(i) = { d N \over dm }(K_i)
\end{equation}
where $dN/dm=\int_0^\infty dz P_f(K,z)$ are the differential number counts. 

\subsection{Probabilities for Cluster Galaxies} \label{sec:clusterprob}

The probability of finding a cluster galaxy of magnitude $K_i$ and redshift
$z_i$ at an angular separation $\theta_{ik}$ from a cluster at redshift
$z_{ck}$ with (proper) scale length $r_{ck}$, galaxy number $N_{*ck}$, and
rest frame velocity dispersion $\sigma_{ck}$ is
\begin{eqnarray}
   P_c(i,k) = N_{*ck} { \phi_c(K_i-{\cal D}(z_{ck})) \over \Phi_c(M_{K*}) } 
   \Sigma\left( { D_A(z_{ck})\theta_{ik} \over r_{ck} }\right)
   \qquad\qquad\nonumber\\
   \quad
   { \exp(-c^2(z_i-z_{ck})^2/2\sigma_{ck}^2(1+z_{ck})^2)\over 
        \sqrt{2\pi}\sigma_{ck}(1+z_{ck}) }. 
   \qquad
\end{eqnarray}
The number of galaxies $N_{*ck}$ is normalized to represent the number of
galaxies brighter than $L_*$ inside the radius $r_{\rm out}\equiv c r_c$. 
This normalization is convenient for calculation, and we discuss 
other normalizations in \S\ref{sec:MassEst}.
The performance of the algorithm in correctly determining
galaxy membership and cluster velocity dispersions is significantly 
enhanced if we optimize the width of the velocity filter for each 
cluster candidate.  We used the range 
$150\hbox{km~s}^{-1} \leq \sigma_{ck} \leq 1200\hbox{km~s}^{-1}$. 
For a galaxy with an unknown redshift,
the likelihood is
\begin{equation}
   P_c(i,k) = N_{*ck} { \phi_c(K_i-{\cal D}(z_{ck})) \over \Phi_c(M_{K*}) }
   \Sigma\left( { D_A(z_{ck})\theta_{ik} \over r_{ck} }\right).
\end{equation}

We calculate the likelihood in $1h^{-1}$~Mpc regions around each galaxy
based on either the measured redshift or the average redshift for galaxies
with the observed magnitude.  The search area was limited to 
$\Delta\theta=4^\circ$.  The fraction of
cluster galaxies inside search radius $R = D_A (z_{ck})\Delta\theta$
of a cluster at redshift $z_{ck}$ is
\begin{equation}
  A_k(\Delta\theta) = { g(R/r_c) \over F(c) } 
           { \Phi_c(K_{\rm lim}-{\cal D}(z_k)) \over \Phi_c(M_{K*}) }.
\end{equation}

\subsection{Likelihood Function}

The likelihood function for finding the first cluster $n_c=k=1$ is
\begin{equation}
   \Delta \ln { \cal L }(k) =  - N_{*ck} A_k + \sum_{j=1}^{N_g} 
       \ln \left( { P_{f}(j) + N_{*ck} P_{ck}(j,k) \over P_{f}(j) } \right) 
\end{equation}
where the sum extends over all galaxies within angle $\Delta\theta$ of the
cluster.  We compute the likelihood at the positions of each survey
galaxy over a range of trial redshifts, optimizing the number of
galaxies $N_{*ck}$ for each trial.  We then add the trial cluster which
produced the largest change in the likelihood to the cluster sample.
We automate the process and avoid finding the same cluster by including
each new cluster in the density model used to find the next cluster,
\begin{equation}
\begin{array}{ll}
   \Delta \ln { \cal L }(n_c) & =  - N_{*c n_c} A_{n_c} \\
   &+ \sum_{j=1}^{N_g}
     \ln \left({ P_{f}(j) +\sum_{k=1}^{n_c} N_{*ck} P_{ck}(j,k) \over 
                 P_{f}(j) +\sum_{k=1}^{n_c-1} N_{*ck} P_{ck}(j,k) } \right).
\end{array}
   \label{eqn:like}
\end{equation}
The clusters $k=1 \cdots n_c-1$ are our current catalog and have fixed
properties, while $k=n_c$ is our new trial cluster.
By this method we ``clean'' the clusters out of the sample in order of
their likelihood (for further discussion see WK).

We found that several small modifications improved the performance of
the algorithm.  In most cases these are priors on the filter variables
$N_{*c}$ and $\sigma_{ck}$ designed to stabilize parameter estimation
for systems with very few galaxies.  First, we added a prior to the
likelihood, Eq.~(\ref{eqn:like}), of the form $-\ln(N_0^2+N_{*c}^2)$ with 
$N_0=0.1$ comparable to the value of $N_{*c}$ expected for a very poor group.
This encodes the information that massive clusters are rare, with a slope
roughly matching the power-law slope of the halo mass function.  In some
senses the prior acts to control the problem of Malmquist biases created
by combining a steep number function $dn/dN_{*c}$ with the uncertainties
in estimates of $N_{*c}$.
Second, when optimizing the likelihood over the width
of the velocity filter $\sigma_{c}$, we included a prior on the 
likelihood of the form $\ln (\sigma_{c}/\sigma_0)$ with the
arbitrary normalization of $\sigma_0=10^3$~km/s.  This prior 
makes the likelihood independent of $\sigma_{c}$
when a candidate cluster contains only one galaxy with a redshift
measurement\footnote{For a maximum likelihood estimate of the
velocity dispersion from a set of velocity measurements $v_i$,
adding this prior leads to an unbiased estimator of the velocity
dispersion.  If $-\ln{\cal L}=N\ln\sigma+\sum_i(v_i-\bar{v})^2/(2\sigma^2)$
adding a $\ln\sigma$ prior and solving for $\sigma$ leads to the unbiased
estimator $\sigma^2=(N-1)^{-1}\sum_i(v_i-\bar{v})^2$.  Without the prior the
estimator is biased by $N/(N-1)$.}. 
Third, we added a weak prior on the relation between $N_{*c}$ and
$\sigma_c$ based on our initial results.  Preliminary, consistent
models found that $\log N_{*c} = 1.13 + 1.90 \log (\sigma_c/1000\hbox{km/s})$
with a dispersion of 0.39~dex.   After determining $N_{*c}$ as a 
function of $\sigma_c$ on a coarse grid, the likelihoods were weighted
by this log-normal prior.  The prior is only important for setting the
filter width $\sigma_c$ in systems with too few galaxies for a direct
estimate of the velocity dispersion.  It also helps to suppress the
superposition of clusters into a single larger cluster because 
superpositions have large apparent velocity dispersions for the
number of galaxies.
Fourth, for the trial galaxy used to compute the likelihood we used the mean
surface density inside the break radius $r_c$ rather than the central density
of the profile in computing the likelihood.
This reduces a bias towards finding clusters centered on isolated galaxies
given our cusped density profile.

\subsection{Matching to Existing Catalogs}

Once we have identified a cluster candidate we want to compare its
properties either to the known properties of our synthetic catalog
or to other properties of the real clusters.  We cross reference 
our cluster catalog to either published or the synthetic cluster
catalogs both through the member galaxies and by matching the
cluster coordinates and redshifts, separately tracking both 
identifications.  Each galaxy is 
associated with its most probable cluster, the cluster which
maximizes its membership likelihood, and has a likelihood 
ratio $\delta_i$ between the probability that it is a member
of that most probable cluster and the probability that it is
a field galaxy.  This likelihood ratio defines the probability
that the galaxy is a cluster member, $p_i = \delta_i/(1+\delta_i)$.  

In matching our new catalog to existing catalogs we require 
algorithms which are fully automated and robust.  They will
not be perfect, but we cannot afford to conduct laborious 
individual case studies in a broad survey.  For  
our synthetic data, each galaxy is labeled by its parent cluster,
allowing us to match the output and input catalogs by identifying 
the most common parent cluster for all the galaxies with membership
probability $p_i > 0.5$ ($\delta_i >1$) in each output catalog.
This means of identification is robust, with relatively few 
multiple identifications or false identifications.

For the real data, the inhomogeneity of local cluster identification 
enormously complicates cross
references between our cluster identifications and previous
studies.  The primary problem is that cross referencing is
best done by cross referencing the proposed member galaxies,
but few local group/cluster surveys supply such data.
We settled on an ecumenical approach using the
NASA Extragalactic Database (NED)
to identify the galaxies likely to be associated
with each known cluster and then matching our cluster
catalog to the existing catalogs based on the member
galaxies.  We searched NED for all ``clusters'' (object types
GClstr, GGroup, GTrpl \& GPair).  From this list of 31130
clusters we looked for overlapping identifications which 
were not recognized by NED.  This required some iteration,
and was a hopeless task at low redshift (near Virgo) where 
a nearly infinite number of ``groups'' have been found,
many of which appear to be substructures of a larger object.
Based on these identifications we assigned cluster
identifications to galaxies within a projected radius
$0.5 h^{-1}$~Mpc and a redshift difference less than $1000$~km/s.
We then attempted to match our cluster detections to these
galaxies using both the match to the cataloged cluster
position and redshift and the most common cluster names
associated with the galaxies we assign to our detection.
The final cross-matching system works reasonably well but
is by no means perfect.  

\subsection{Mass Estimates} \label{sec:MassEst}

In addition to identifying the parent cluster and the member
galaxies we would also like to estimate the mass of the cluster.
We have four different ways of making the comparison.
First, for the synthetic catalogs we have direct estimates of the masses.
Second, for the real data we have measurements of X-ray luminosities and
temperatures which may serve as surrogate estimates of the mass.
Third, we have the characteristic number of galaxies in the clusters.
Fourth, we have estimates of the galaxy velocity dispersion in the
clusters.

Because clusters lack sharp edges, we must exercise some care in defining
and comparing our mass estimates (White~\cite{HaloMass},\cite{MassFunction}).
In fact, we found that {\it failure to properly match mass definitions affects
both the normalization and the slope of our estimates of the halo multiplicity
function}.  For this reason we give explicit details of our procedure below.

We use two standard mass estimates for the clusters in our
synthetic catalog.  The friends-of-friends mass, $M_{FoF}$,
is the total mass of the N-body particles linked into a 
group using the FoF algorithm with a linking length of
$b=0.2$ in units of the mean inter-particle spacing.  We 
assign galaxies to the clusters based on the FoF mass (Eq.~\ref{eqn:nfof})
because it can be computed quickly while we are building the 
synthetic galaxy catalogs.  We also calculate separately the 
mass $M_{200}$ inside the radius $r_{M200}$ for which the enclosed 
density is $\Delta_M=200$ times the {\it critical density}.  When
we compare these two estimates we find that $M_{200}\simeq 0.66 M_{FoF}$ 
because the FoF mass corresponds to a lower density threshold of
$\Delta_M \simeq \Omega_M/b^3 \sim 40$.  
  
For the real data, where we do not measure the masses directly,
we will compare our estimates from the galaxies to X-ray 
luminosities and temperatures.
We used compilations of X-ray luminosities from 
B\"ohringer et al.~(\cite{Bohringer00}),
Cruddace et al.~(\cite{Cruddace01}),
Ebeling et al.~(\cite{Ebe96}, \cite{Ebe98}, \cite{Ebeling00}),
De Grandi et al.~(\cite{deGrandi99}),
Mahdavi et al.~(\cite{Mahdavi00}),
and Reiprich \& B\"ohringer~(\cite{Reiprich02}), 
and compilations of X-ray temperatures from 
David et al.~(\cite{David93}),
Ponman et al.~(\cite{Ponman96}),
Markevitch~(\cite{Mar98}),
Helsdon \& Ponman~(\cite{Helsdon00}),
Ikebe et al.~(\cite{Ikebe02}) and
White~(\cite{dwhite00}).
Most of the X-ray luminosity data is drawn from the ROSAT All-Sky
Survey (RASS, Voges et al.~\cite{Voges99}).
We rescaled the X-ray luminosities and fluxes to a common scale
by allowing for logarithmic offsets relative to Reiprich \& 
B\"ohringer~(\cite{Reiprich02}) for each survey, with the
offsets determined from clusters with measurements in multiple
surveys.  The offsets we applied are listed in Table~\ref{tab:offsets}.
Measured scaling relations for clusters of galaxies are
(Markevitch~\cite{Mar98})
\begin{equation}
   \log\left({ L_X h^2\over 10^{44}\hbox{ergs/s}}\right) = (0.15\pm0.04)
        + (2.10\pm0.24) \log \left( { T\over 6\hbox{keV}}\right)
  \label{eqn:LxTx} 
\end{equation}
and (Horner, Mushotsky \& Scharf~\cite{HorMusSch};
Nevalainen, Markevitch \& Forman~\cite{NevMarFor};
Finoguenov, Reiprich \& B\"ohringer~\cite{FinReiBoh};
Xu, Jin \& Wu~\cite{XJW01})
\begin{equation}
  M_\Delta \simeq
      10^{15} h^{-1}M_\odot\ \left({T_X\over 1.3{\rm keV}} \right)^{3/2}
      \left( \Delta_M E^2 \right)^{-1/2}
  \label{eqn:MTx} 
\end{equation}
where $\Delta_M$ is the density threshold used to estimate the mass
$M_\Delta$,
and $E(z)\equiv H(z)/H_0=1$ for our low redshift sample.
The normalization of this last relation is currently quite uncertain, at the
30-50\% level, due to the notorious difficulties inherent in measuring the
mass of a cluster (e.g.~Evrard, Metzler \& Navarro~\cite{EMN96})
and differences in methods for fitting temperatures to the X-ray emitting ICM.
(See Fig.~2 of Huterer \& White~\cite{HutWhi} for a compilation of theoretical
and observational results.)
Different authors use different methods to model the `temperature' of
the plasma, different methods of estimating the mass and indeed even
different {\it definitions\/} of the mass!
To cloud the picture even further, numerical simulations based on purely
adiabatic gas physics find a different normalization of this relation than
most of the observations (although it must be noted that significant progress
in this regard has occurred recently,
e.g.~Muanwong et al.~\cite{MTKP}, Voit et al.~\cite{Voit02}).

\begin{deluxetable}{crrr}
\tablecaption{Adopted Offsets Between X-ray Data}
\tablewidth{0pt}
\tablehead{ 
Name      &  Offset   &    Error   & $N_{\rm cluster}$ }
\startdata
BCS       & $ 0.036  $  & $ 0.005   $  & 351 \nl
de Grandi & $ 0.246  $  & $ 0.009   $  & 145 \nl
Cruddace  & $ 0.039  $  & $ 0.008   $  & 109 \nl
Ponman    & $-0.084  $  & $ 0.035   $  &   9 \nl
RASSCALS  & $ 0.067  $  & $ 0.008   $  &  55 \nl
HIFLUGCS  & $\equiv 0$  & $ \equiv 0$  & 193 \nl
XBACS     & $ 0.040  $  & $ 0.004   $  & 377 \nl
NORAS     & $ 0.051  $  & $ 0.004   $  & 299 \nl
\hline
\enddata
\tablecomments{The offsets, defined by the logarithmic correction needed
to match the fluxes quoted by
Reiprich \& B\"ohringer~(\protect{\cite{Reiprich02}}),
applied to the X-ray luminosities from the different samples we analyzed.
The offsets were generally small, with only that applied to the sample of
de Grandi et al.~(\protect{\cite{deGrandi99}}) being significant.}
\label{tab:offsets}
\end{deluxetable}

\begin{deluxetable}{llllccl}
\tablecaption{Estimates of Halo Occupancy Function}
\tablewidth{0pt}
\tablehead{ 
   Data     &Mass Scale &Correlation &Transformations &zero-point $A$   &slope $B$ &Comments 
    }
\startdata
Synthetic &True                            
          &                         
          &                         
                                                             &$1.32$           &$0.98$          \nl
          &$M_{200}$
          &$N_{*666}(M_{200})$           
          &                 
                                                             &$1.25\pm0.03$    &$0.89\pm0.02$   \nl
          &$\sigma_{200}(M_{200})$, Eq.~\protect{\ref{eqn:Msigma}}            
          &$N_{*666}(\sigma)$            
          &                 
                                                             &$1.23\pm0.04$    &$0.94\pm0.03$   \nl
          &$dn/dM_{200}$                  
          &$dn/dN_{*666}$                  
          &
                                                             &$1.43\pm0.05$    &$1.13\pm0.04$   &$N_{thresh}=3$\nl
          &
          &
          &
                                                             &$1.37\pm0.05$    &$1.04\pm0.04$   &$N_{thresh}=5$\nl
          &
          &
          &
                                                             &$1.43\pm0.09$    &$1.09\pm0.08$   &$N_{thresh}=10$\nl
\hline
2MASS     
          &$\sigma_{200}(M_{200})$, Eq.~\protect{\ref{eqn:Msigma}}            
          &$N_{*666}(\sigma)$            
          &
                                                             &$1.28\pm0.03$    &$0.90\pm0.02$   \nl
          &
          &$N_{*666}(T_X)$               
          &$\sigma(T_X)$, Eq.~\protect{\ref{eqn:sigmaT}}
                                                             &$1.24\pm0.05$    &$1.13\pm0.09$  &Bright X\nl
          &
          &
          &
                                                             &$1.25\pm0.05$    &$1.06\pm0.06$  &All X\nl
          &
          &$N_{*666}(L_X)$               
          &$\sigma(L_X)$, Eq.~\protect{\ref{eqn:sigmaL}}
                                                             &$1.22\pm0.04$    &$0.98\pm0.06$  &Bright X\nl
          &
          &Average
          &
                                                             &$1.25\pm0.03$    &$1.02\pm0.10$  &  \nl
          &$T_X(M_{200})$, Eq.~\protect{\ref{eqn:MTx}}               
          &$N_{*666}(\sigma)$               
          &$\sigma(T_X)$, Eq.~\protect{\ref{eqn:sigmaT}}
                                                             &$1.57\pm0.04$    &$1.10\pm0.02$  &\nl
          &
          &$N_{*666}(T_X)$               
          &
                                                             &$1.60\pm0.07$    &$1.39\pm0.11$  &Bright X\nl
          &
          &
          &
                                                             &$1.59\pm0.06$    &$1.31\pm0.08$  &All X\nl
          &
          &$N_{*666}(L_X)$               
          &$L_X(T_X)$, Eqs.~\protect{\ref{eqn:LxTx}}
                                                             &$1.56\pm0.06$    &$1.05\pm0.07$  &Bright X\nl
          &
          &
          &$L_X(T_X)$, Eqs.~\protect{\ref{eqn:LxTx2}}
                                                             &$1.56\pm0.06$    &$1.19\pm0.08$  &Bright X \nl
          &
          &Average
          &
                                                             &$1.58\pm0.02$    &$1.21\pm0.14$  &  \nl
          &$dn/dM_{200}$                  
          &$dn/dN_{*666}$  
          &
                                                             &$1.49\pm0.04$    &$1.13\pm0.03$  &$N_{thresh}=3$ \nl
          &
          &
          &
                                                             &$1.49\pm0.04$    &$1.08\pm0.03$  &$N_{thresh}=5$ \nl
          &
          &
          &
                                                             &$1.54\pm0.05$    &$1.13\pm0.06$  &$N_{thresh}=10$ \nl
          &              
          &Adopted Standard
          &              
                                                             &$1.44\pm0.17$    &$1.10\pm0.09$  &\nl
\hline
\enddata
\tablecomments{The halo occupancy function has the form
  $\log N_{*666} = A + B \log (M_{200} h/10^{15}M_\odot)$.  Except for the fits to the 
  cluster number function, all results are for a standard cluster sample consisting of 
  the systems with at least $N_v\geq 5$ associated redshifts.  The cluster number function
  estimates are for threshold galaxy numbers of $N_{thresh}=3$, $5$ and $10$.
  The halo mass function $dn/dM_{200}$ is that of Jenkins et al.~(\protect\cite{JFWCCEY}) converted
  to $M_{200}$ from the original FoF masses assuming $M_{200}=0.66M_{FoF}$ as found in our
  simulations for the same FoF linking length.  The N-body simulations used to generate
  our synthetic catalog have the same initial parameters.  Models based on X-ray data
  samples limited by $T_X\geq 1$~keV and $L_X \geq 10^{42}h^2$~ergs/s contain the
  comment ``Bright X''.  The Average entries are simply the average coefficients and
  their scatter.  The Adopted standard combines the 2 Average relations and the
  $N_{thresh}=5$ result for the 2MASS data.
  }
\label{tab:multiplicity}
\end{deluxetable}

The number of member galaxies or their aggregate luminosity must
be carefully defined before we can compare it to other results
or theoretical models.  Our fit parameter $N_{*c}$ is defined to be
the number of $L \geq L_*$ galaxies inside the fixed, spherical
outer radius $r_{\rm out}= c r_c = 0.8h^{-1}$~Mpc.  Postman et al.~(\cite{MF},
\cite{Postman02}) and Donahue et al.~(\cite{Donahue01})
use the luminosity, measured in units of $L_*$, rather than the
number. We prefer the number of galaxies, $N_*$, to this equivalent 
luminosity, $\Lambda_*$, because it has a closer relation to the Poisson 
statistics which determine the detectability of clusters and may
depend less on the defining luminosity function or wavelength.
For the same normalizing aperture, the two quantities are related by 
\begin{equation}
   \Lambda_* L_*=L_* N_{*} \Gamma[2+\alpha]/\Gamma[1+\alpha,1]= 5.0 N_* L_*
   \label{eqn:postman}
\end{equation}
for $\alpha=-1.09$.  We cannot precisely match the Postman et al.~(\cite{MF}) 
definition of $\Lambda_{cl}$ because we use different filters and
luminosity functions.  To the extent that $\Lambda_{cl}$ is defined
by a $r_{cl}=1.1h^{-1}$~Mpc ($1.5h_{75}^{-1}$~Mpc) radius aperture, we
expect $N_{*cl}\simeq(g(r_{cl}/r_c)/F(c))N_{*c}=1.6 N_{*c}$.

While $N_{*c}$ and $\Lambda_{cl}$ are natural variables for finding 
clusters, they are poor choices for comparing to theoretical models.  
For comparisons to theoretical models we will use the number of $L>L_*$
galaxies $N_{*\Delta}$ inside the spherical radius $r_{N\Delta}$ such that
the enclosed galaxy density is $\Delta_N$ times the {\it average density\/}
of galaxies.
Given a cluster with $N_{*c}$ galaxies brighter than $L_{*}$ inside
$r_{\rm out}$ we determine $r_\Delta$ and $N_{*\Delta}$ by solving the defining
relations
\begin{equation}
  N_{*\Delta} \equiv N_{*c} F(r_{N\Delta}/r_c)/F(c) = 
   { 4 \pi \over 3 } n_* \Delta r_{N\Delta}^3\Gamma[1+\alpha,1].
\end{equation}
For reasons of computational efficiency and stability we keep $c$ fixed
when we determine $r_{N\Delta}$.  In most theoretical models, the 
distribution of galaxies basically traces the distribution of mass
(Katz, Hernquist \& Weinberg~\cite{KatHerWei};
 Gardner et al.~\cite{GKHW};
 Pearce et al.~\cite{Pearce};
 White, Hernquist \& Springel~\cite{WhiHerSpr}).
However, if we assume the galaxies are unbiased\footnote{The `galaxies' in the
simulated catalog are very nearly unbiased.  The bias of the 2MASS galaxies is
currently unknown, but we shall argue later it is likely they are relatively
unbiased.},
to isolate the same physical region we must set
$\Delta_N = \Delta_M/\Omega_M$ because the galaxy overdensity is defined
relative to the average galaxy density while the mass overdensity is defined
relative to the critical density rather than the mean mass density.
Thus, taking $\Omega_M=0.3$, the number of galaxies brighter than $L_{*}$
inside the region with the standard mass overdensity of $\Delta_{M}=200$ is
$N_{*666}=0.66 N_{*FoF}$ with $r_{M200}=r_{N666}$.

Our best theoretical estimate of the conversion to Postman et al.~(\cite{MF}, \cite{Postman02})
and Donahue et al.~(\cite{Donahue01}) is 
\begin{equation}
  \Lambda_{cl} \simeq 11 N_{*666}^{3/4}
   \label{eqn:postman2}
\end{equation}
for $\Lambda_{cl}\gtorder 10$.  The slope difference is created by the difference
between a fixed $r_{cl}=1.5h_{75}^{-1}$~Mpc aperture for $\Lambda_{cl}$ and
a variable aperture $r_{N666}$ for $N_{*666}$.  In practice, both the normalization 
and slope could differ from this estimate because of differences in the luminosity
functions, wavelengths (K band versus I band) and matched filter structures. A
simple check is to compare $\Lambda_{cl}$ and $N_{*666}$ values for clusters
of different Abell~(\cite{Abe58}) cluster richness classes. 
Our overlap with the Abell~(\cite{Abe58}) catalog is limited because the
average redshift of the Abell clusters is significantly higher than our
2MASS catalog.  A quick match based on NED cluster names found 31, 39, 
and 15 Abell richness class 0, 1 and 2 clusters with fractional errors 
in $N_{*666}$ smaller than 25\% in the catalogs of \S\ref{sec:clusters}.
We find an average richness and dispersion of
$\langle N_{*666}\rangle = 5.3\pm 2.7$, $8.9\pm 4.3$ and $13.4\pm6.4$ for
Abell richness classes of 0, 1 and 2 respectively.
The ratios of the average $N_{*666}$ values roughly track the ratios of the
number of galaxies associated with each richness class.
Postman et al.~(\cite{Postman02}) estimate that $\Lambda_{cl}=50$, $80$ and
$130$ should correspond to the typical Abell richness classes 0, 1 and 2
clusters which differs significantly from our estimated conversion to
$\Lambda_{cl}$ (Eq.~\ref{eqn:postman2}) of $\Lambda_{cl}=41\pm16$, $60\pm22$, 
and $84\pm28$ for the same richness classes.  While using Abell richness classes 
to compare modern algorithms can hardly be recommended, the comparison suggests
that we should explore empirical conversions between $\Lambda_{cl}$ and 
$N_{*666}$.

We find clusters with richnesses as low as $N_{*666}\simeq 0.1$ in our
catalogs, and this requires some explanation.  First, low values of
$N_{*666}$ only imply that it is unlikely the cluster contains any 
bright ($L > L_*$) galaxies.  It does not imply the cluster is unlikely
to contain galaxies.  For example, an $N_{*666}=0.3$ cluster should 
contain 3 (7) galaxies with $L>L_*/10$ ($L>L_*/100$) inside its 
virial radius.  We can routinely detect such systems provided the
redshift is low enough for us to detect the lower luminosity 
galaxies. Second, we detect the systems in projection,
and the number of galaxies projected inside the virial 
radius is larger than the number inside the spherical virial radius.
This is a modest effect for the NFW profile, but real clusters are
embedded in more extended infall regions which boost projection
effects.  Third, there are significant errors in these low estimates 
of $N_{*666}$ because there are so few galaxies.  Thus, these low
richness clusters can be real virialized systems, but they must 
be treated with care.  

Finally, we have estimates of the velocity dispersion of the
clusters.  Although we optimized the cluster velocity width
$\sigma_{ck}$ as part of our matched filter, we do so at low
resolution (150~km/s) and it includes contributions from 
galaxies at large distances from the cluster center.  To
better mimic a virial velocity for a cluster we use a
membership probability weighted estimate of the velocity
dispersion.  We take all galaxies with measured redshifts,
membership probabilities $p_i \geq 0.5$ and projected separations
less than the model outer radius $R \leq c r_c = 0.8h^{-1}$~Mpc
and estimate a mean redshift
\begin{equation}
    \langle c z \rangle = \sum c z_i p_i / \sum p_i
\end{equation}
and a velocity dispersion
\begin{equation}
     \sigma_0^2 = { (\sum p_i)^2  \over (\sum p_i)^2  - \sum p_i^2 } 
     \  { \sum ( c z_i - \langle c z \rangle )^2 p_i \over  \sum p_i }
     \label{eqn:veldisp}
\end{equation}
which we then correct to the rest frame $\sigma=\sigma_0/(1+z)$.  
The leading term plays the role of the familiar $N/(N-1)$ factor
needed to make the variance of the velocities about the mean an
unbiased estimator of the velocity dispersion. The
weighting by membership probability plays an equivalent role to the
clipping of outliers in standard approaches to estimating cluster
velocity dispersions (see discussion in Borgani et al.~\cite{BGCYE}).
and the method is similar to that
used by Barmby \& Huchra~(\cite{Barmby98}).

\begin{figure}
\begin{center}
\resizebox{3.3in}{!}{\includegraphics{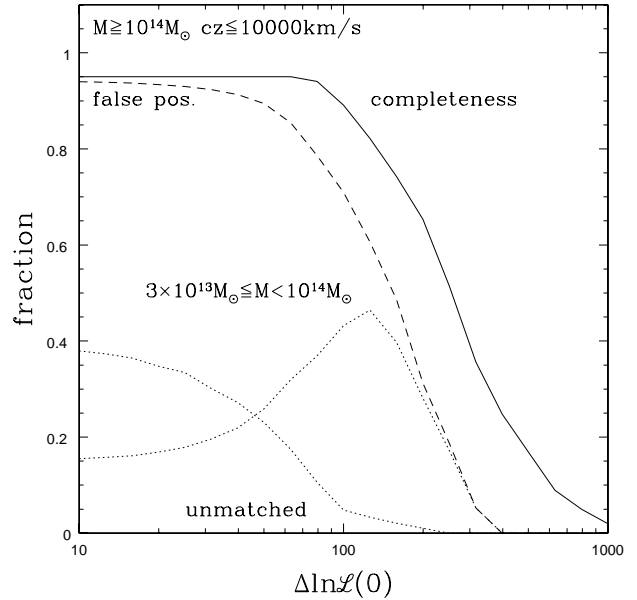}}
\end{center}
\caption{\footnotesize%
Completeness as a function of $\Delta\ln{\cal L}(0)$ for clusters with
$M_{\rm cut}\geq 10^{14}h^{-1}M_\odot$ and $cz \leq 10000$~km/s.
The solid line shows the completeness, the dashed line shows the fraction of
cluster candidates which are false positives.
The upper dotted line (at $\Delta\ln{\cal L}(0)=100$)
shows the fraction of cluster candidates which are real
clusters with masses $M_{\rm cut}/3 \leq M < M_{\rm cut}$.
The lower dotted line (at $\Delta\ln{\cal L}(0)=100$)
shows the fraction of cluster candidates which cannot be
identified with any halo.}
\label{fig:comp1}
\end{figure}

\begin{figure}
\begin{center}
\resizebox{3.3in}{!}{\includegraphics{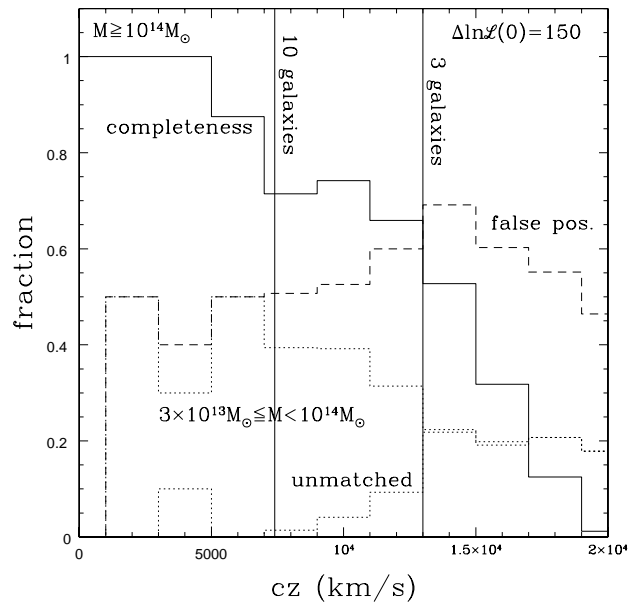}}
\end{center}
\caption{\footnotesize%
Completeness as a function of redshift for clusters with
$M_{\rm cut}\geq 10^{14}h^{-1}M_\odot$ and $\Delta\ln{\cal L}(0)=150$.
The solid histogram shows the fraction of the $M\geq M_{\rm cut}$ clusters
detected in each bin, the dashed histogram the fraction of candidates which
are false positives.
The dotted histogram, decreasing with redshift, shows the
fraction of cluster candidates which are real clusters with
$M_{\rm cut}/3 \leq M < M_{\rm cut}$, while the rising dotted histogram shows
the fraction of candidates which cannot be identified with any halo.
The vertical lines mark the redshifts where the average $M=M_{\rm cut}$
cluster has $10$ or $3$ galaxies.}
\label{fig:comp2}
\end{figure}

\subsection{The Cluster Number Function} \label{sec:dndN}

We can also estimate the cluster number function, $dn/dN$, which is the 
distribution of clusters with respect to the number of member galaxies.
If the number of member galaxies, $N(M)$, is a simple function of the cluster 
mass then the number function is simply related to the cluster mass
function $dn/dM$ by
\begin{equation}
  {dn\over dN} = {dn\over dM}\left| {dM\over dN}\right|.
  \label{eqn:dndNanal}
\end{equation}
If we assume that the cluster mass function is known, then we can infer
the halo occupancy function by determining the function $N(M)$ needed
to transform the assumed mass function into the observed number function.
We use the Jenkins et al.~(\cite{JFWCCEY}) CDM mass function, which
should closely match that of our N-body simulation since it is based on
numerical simulations with similar cosmological parameters
(see White~\cite{MassFunction} for further discussion).
The Jenkins et al.~(\cite{JFWCCEY}) mass function was derived using $M_{FoF}$
masses for the clusters, so we transformed it from $dn/dM_{FoF}$ to
$dn/dM_{200}$ assuming the average relation that $M_{200}=0.66M_{FoF}$ we
observe in our simulations for the same linking length.
This method is similar to the approach used by
Berlind \& Weinberg~(\cite{Berlind02}) and Berlind et al.~(\cite{BerSDSS}).

We can estimate the number function using the $V/V_{\rm max}$ method
(e.g.~Moore et al.~\cite{MooFreWhi}) as follows.
At redshift $z$, a cluster normalized by $N_{*666}$ contains
\begin{equation}
  N_g(z)=N_{*666} \Gamma[1+\alpha,L_{\rm lim}(z)/L_*]/\Gamma[1+\alpha,1]
\end{equation} 
galaxies, where $L_{\rm lim}(z)$ is the luminosity corresponding to
the survey magnitude limit at redshift $z$.
Suppose we detect all clusters with at least $N_g \geq N_{\rm thresh}$ galaxies.
Then we detect clusters in the volume $V_{\rm max}(N_{*666})$ defined by the
comoving survey volume out to the redshift $z_{\rm lim}$ which solves
$N_{\rm thresh}=N_g(z_{\rm lim})$.
The integral number function is then
\begin{equation}
    n (>N) = \sum_{{\cal N}\geq N} V_{\rm max}({\cal N})^{-1}
\end{equation}
where the sum includes only clusters with more than the threshold
galaxy number at their observed redshift.
For large values of $N_{\rm thresh}$ we are less sensitive to the
Poisson errors in estimating $N_{*666}$ and minimize the risk of 
including false positives as clusters.
This comes at the price of increased Poisson errors in the number function,
because we include fewer clusters, and increasing sample variance because we
include less volume.   We limited the sample used and $V_{\rm max}$ to
$cz < 15000$~km/s and used bootstrap resampling of the catalog to
estimate the uncertainties in the number function.

\section{Tests With the Synthetic Catalog} \label{sec:tests}

We start by testing the algorithm and our recovery of the halo 
occupancy function in the synthetic data designed to mimic the
real 2MASS galaxy sample. 

\subsection{Completeness and contamination}

\begin{figure}
\begin{center}
\resizebox{3.3in}{!}{\includegraphics{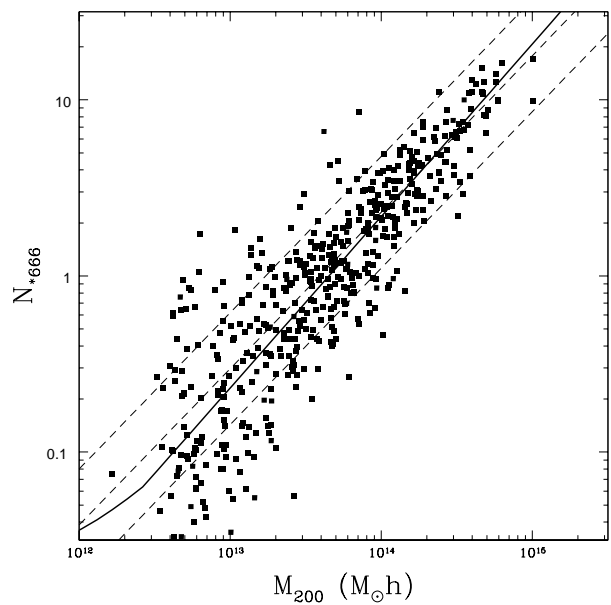}}
\end{center}
\caption{\footnotesize%
The number of galaxies $N_{*666}$ as a function of the virial mass $M_{200}$.
The points show all detected clusters with at least $N_v\geq 5$ associated
redshifts which have been matched to the input cluster catalog. The heavy solid
line shows the true relationship and the light dashed lines show the estimated
relation and its width as estimated from the variance of the points. } 
\label{fig:nstartest}
\end{figure}

We first consider the completeness of the resulting catalogs.  As we discuss
in WK, the likelihood associated with a cluster roughly scales with the number
of member galaxies, so the distance at which we can detect clusters can be
roughly estimated from the number of galaxies.
In our models, the average number of cluster galaxies drops to $N_{FoF}=10$
($3$) at redshifts of $cz=12000$ ($17000$), $7400$ ($13000$),
and $2700$ ($7500$)~km/s for clusters of mass $M h/M_\odot=3 \times 10^{14}$,
$10^{14}$ and $3 \times 10^{13}$ respectively.     
The spread in the likelihoods for a fixed mass and redshift will be
considerable because of the Poisson variations in the number of galaxies
bright enough to be visible and because of the random variations in the
numbers of measured redshifts for the fainter galaxies.
In WK we showed that the cluster likelihood
essentially scales with the number of member galaxies, so that by scaling the
selection likelihood as $\Delta\ln {\cal L}(z)=C(z)\Delta\ln {\cal L}(0)$ we
can extract clusters with a nearly constant mass threshold as a function 
of redshift.  The choice of the zero-redshift likelihood
threshold, $\Delta\ln {\cal L}(0)$, is related to the mass cut
and its completeness.  

Fig.~\ref{fig:comp1} illustrates this for model clusters with
$M \geq 10^{14}h^{-1}M_\odot$.  
Of the 101 clusters with $M \geq 10^{14}h^{-1}M_\odot$ and $cz < 10^4$~km/s,
we miss only 5 at our likelihood cutoff.  Of these, one was merged into
another cluster, 4 were systems with $M\simeq 10^{14}h^{-1}M_\odot$
and fewer galaxies than expected given their mass. 
We can achieve very high completeness out to the redshift where the typical
cluster contains only 3 galaxies.  However, all samples with high
completeness also have high false positive fractions.  If the likelihood
threshold is kept high enough, the vast majority of these false positives are 
real clusters with masses just below the cutoff mass scale -- very few of these 
false positives lack identification with any $M >3\times 10^{13}h^{-1}M_\odot$ halo.
For lower likelihood thresholds the mass range of the false positives broadens
and then slowly becomes dominated by unmatched systems where we could not 
identify the detection with a cluster in the input catalog.

Fig.~\ref{fig:comp2} shows the redshift dependence of the completeness and
false positive rates after choosing $\Delta\ln {\cal L}(0)$ to have a 50\%
false positive rate.  The completeness declines slowly with redshift, but
remains close to 90\% complete from $0 \leq cz \leq 10000$~km/s.
The false positive rate averages 50\%, and the false positives consist
almost entirely of real clusters with masses between
$3\times 10^{13}h^{-1}M_\odot$ and $10^{14}h^{-1}M_\odot$.
False positives that correspond to no cluster become important only for redshifts
beyond which a $10^{14}h^{-1}M_\odot$ cluster contains fewer than 3 galaxies.
Within our $cz \leq 10000$~km/s selection window there is a total volume of
$4.1 V_0$ where $V_0 = (100/h)^3$~Mpc$^3$, containing 101 clusters more
massive than $10^{14}h^{-1}M_\odot$.
Using these clusters to estimate the mass function would be limited by the
correction for the contamination by lower mass clusters rather than the
correction for completeness or statistical uncertainties including sample
variance.  

The true completeness of the recovered cluster sample is higher than we have
quoted because of the way we have generated the synthetic catalog.
In regions of high density, where we expect the most massive clusters,
the FoF algorithm with a linking length of 0.2 is prone to merging sub-clumps
into larger objects.
Our search algorithm instead detects the individual sub-clumps with lower
likelihoods than would be expected for an object as massive as the FoF cluster.
The sub-clumps may have too low a likelihood to be included (creating a false
impression of incompleteness) or multiple clumps may be included (adding to
the false positives).  Most peculiar high FoF mass but low detection likelihood
systems could be traced to this feature of the FoF algorithm.  

\subsection{Halo occupancy function} \label{sec:hof}

\begin{figure}
\begin{center}
\resizebox{3.3in}{!}{\includegraphics{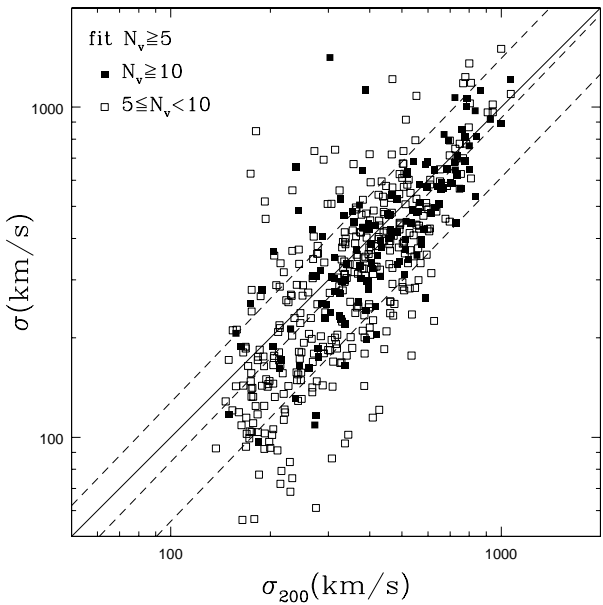}}
\end{center}
\caption{\footnotesize%
Comparison of the velocity dispersion $\sigma$ estimated for the cluster
candidate to the velocity dispersion $\sigma_{200}$ of the dark matter 
particles inside $r_{200}$.
Cluster candidates including at least $N_v\geq 10$ ($5\leq N_v <10$)
redshifts are shown as the solid (open) points.
The solid line indicates $\sigma=\sigma_{200}$ while the
dashed line is the best-fit power law.
The overall correlation is excellent, but there are outliers due to
contamination. }
\label{fig:sigtest}
\end{figure}

Given a well-defined cluster catalog, the next step is to examine how well the
cluster parameters agree with the true properties of the clusters.  We first
test our ability to recover the true halo occupancy function $N(M)$ given
our knowledge of the halo masses for the synthetic catalogs.  For the 
real 2MASS data we will consider cluster X-ray luminosities and temperatures
or galaxy velocity dispersions as surrogate, but independent, estimates for
the mass.

\begin{figure}
\begin{center}
\resizebox{3.3in}{!}{\includegraphics{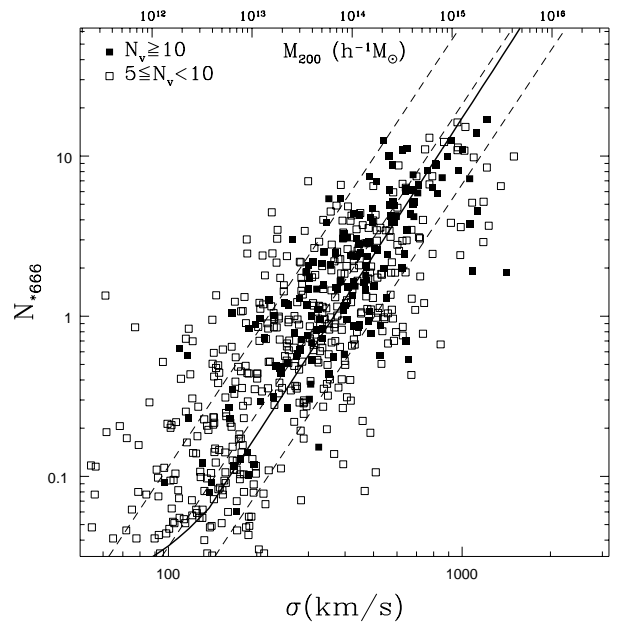}}
\end{center}
\caption{\footnotesize%
The number of galaxies $N_{*666}$ as a function of the estimated cluster
velocity dispersion $\sigma$ for the synthetic data.  Filled (open) points
have $N_v\geq 10$ or ($5 \leq N_v < 10$) velocities contributing to the
dispersion estimate.  The heavy solid line shows the true scaling, and the
dashed lines show the best fit power law and its width as estimated from
the variance of the points.}
\label{fig:numsig}
\end{figure}

We test our ability to recover the halo occupancy function by comparing
the number of galaxies $N_{*666}$ inside the $\Delta_N=666$ contrast level.
This corresponds to the number of galaxies inside a mass contrast level
of $\Delta_M=200$ and should, on average, match $0.66 N_{*FoF}$. There
are two critical issues to estimates of the halo occupancy function.

First, definitions are critical!  The ``number of galaxies in a cluster'' 
must be precisely and similarly defined both for the algorithm used
to find the clusters and the theoretical model used to interpret the
results.  Failure to match the definitions, for example by using $N_{*c}$ 
instead of $N_{*666}$, makes it impossible test the algorithm
(for our synthetic catalog) or to interpret the results
(for the 2MASS catalog).
In particular, $N_{*c} > N_{*666}$ for low mass clusters where
$r_{M200}<c r_c=0.8h^{-1}$~Mpc and $N_{*c} < N_{*666}$ for high mass clusters
where $r_{M200}>c r_c=0.8h^{-1}$~Mpc.
Thus, the slope of $N_{*c}(M)\sim M^{0.75}$ is significantly flatter than
the slope of $N_{*666} \sim M$.
We expect the same flattening of the slope will be found for $\Lambda_{cl}$
as used by Postman et al.~(\cite{MF}, \cite{Postman02}) and Donahue et al.~(\cite{Donahue01})
for similar reasons.

Second, in estimating relations like $N(M)$ it is important to select
subsamples which minimize the introduction of mass and number-dependent
biases.
For example, estimating $N(M)$ using a sub-sample selected such that the
standard error in $N$ is less than a threshold or that the number of galaxies
in the cluster exceeds some threshold will preferentially include low mass
clusters with high values of $N$ over those with low values of $N$.
This biases estimates of $N(M)$ to have flatter slopes than the true relation.

With the logarithmic prior for $N_{*c}$ included in the likelihood, our
ability to estimate the halo occupancy becomes very robust.  Simple 
redshift cuts work well, but we will define our standard sample by
systems which contained at least $N_v \geq 5$ galaxies with redshifts.   
 After selecting a
cluster sample we fit the distribution as a power law including the
formal uncertainties in the estimate of $N_{*666}$ for each cluster.
The results, illustrated in Fig.~\ref{fig:nstartest}, are remarkably good.  Above 
$10^{13}h^{-1}M_\odot$ the input relation is well approximated by a power 
law (see Table~\ref{tab:multiplicity}), 
$\log N_{*666}= A+B\log(M_{200}h/10^{15}M_\odot)$ with 
$A=1.32$ and $B=0.98$.  With our standard sample we find
$A=1.25\pm0.03$ and $B=0.89\pm0.02$.  The error bars are small because
of the large number of clusters (484), 
so the fits formally disagree with the input relation by 3--4 standard
deviations.  

The uncertainties in the halo occupancy function are dominated by systematic
errors arising from sample selection rather than statistical errors.
For example, if we simply include all systems
with $cz \leq 15000$~km/s, roughly the limit for detecting $L_*$
galaxies, we find parameters ($A=1.20\pm0.02$ and $B=0.92\pm0.02$) 
which has a lower normalization, a steeper slope and more scatter
about the mean relation.  For systems with $cz \leq 10000$~km/s,
we find $A=1.23\pm 0.02$ and $B=0.95\pm0.01$.  If we use
a likelihood-dependent redshift cut such as  
$cz < 10000\log(\ln{\cal L})$~km/s, we find $A=1.26\pm0.02$ and $B=0.95\pm0.01$.
We will adopt the $N_v\geq 5$ sample as our standard from this point,
and summarize the estimates of the occupancy function in 
Table~\ref{tab:multiplicity}.

In summary, we can recover the true halo occupancy function of the
synthetic catalog reasonably well.  The uncertainties are dominated
by systematic errors at the level of 0.05 in the slope and 10\%
in the normalization of the relation.  Some of these problems
are due to the generation of the synthetic catalog rather than
the search for clusters in the synthetic catalog.  The FoF 
algorithm used to find clusters in the N-body simulation 
is prone to linking objects which both the eye and our matched
filter will regard as neighboring but separate objects.  These
then appear in our test as objects with high `true' mass but low
`recovered' mass, possibly creating a modest bias towards flatter slopes
and lower normalizations.  While this aspect could be improved upon, we 
will leave exploration of `unlinking' algorithms to future work.

Unfortunately in the real data we do not know the intrinsic masses of
the halos and must use a surrogate estimate such as the cluster 
velocity dispersion, X-ray luminosity or X-ray temperature.  
The surrogate we estimate
as part of our algorithm is the cluster velocity dispersion $\sigma$,
and in Fig.~\ref{fig:sigtest} we compare our estimated velocity dispersions 
$\sigma$ based on Eq.~(\ref{eqn:veldisp}) to the velocity dispersion of 
the dark matter particles $\sigma_{200,DM}$ inside $r_{M200}$ of the 
cluster center.  We select
the sample for comparison with both our standard limits on the
redshift ($cz<15000$~km/s) and a minimum number $N_v\geq 5$ of velocities 
to be included in the redshift estimate.  Fit as a power law
with $(\sigma/10^3\hbox{km/s}) = a(\sigma_{200,DM}/10^3\hbox{km/s})^b$
we find little offset, $a=0.93\pm0.03$, or slope difference,
$b=1.04\pm0.04$.  Restricting the
analysis to systems with $N_v \geq 10$ produces similar results,
with $a=0.89\pm0.05$ and $b=0.94\pm 0.06$.  The scatter is consistent
with the uncertainties in the velocity dispersion estimates.

We can also estimate $N_*(\sigma)$, as shown in Fig.~\ref{fig:numsig}.
In the simulations the velocity dispersion is related to the mass by
\begin{equation}
  \sigma_{200}\simeq 925\left( { M_{200} h \over 10^{15}M_\odot }\right)^{0.34}
  ~\hbox{km/s}
  \label{eqn:Msigma}
\end{equation}
with a 10\% dispersion about the power-law.  The power law describing the
input $N_{*666}(\sigma_{200})$ relation has $N_0=25.6$ and $x=2.75$
for $N_{*666}=N_0(\sigma/10^3\hbox{km/s})^x$.
The observed relation, with our standard selection procedures, is
nearly identical, with $x=2.76\pm0.08$ and $N_0=21.1\pm 2.0$, to
the input relation. The
uncertainties are large because both quantities have relatively
large statistical uncertainties.  These parameters change little
as we adjust the sample selection criteria.  
Alternatively, we can use Eq.~(\ref{eqn:Msigma})
to transform from $\sigma$ to $M_{200}$ to estimate the occupation
number directly, with results very close to the input relation
(see Table~\ref{tab:multiplicity}).  

\begin{figure}
\begin{center}
\resizebox{3.3in}{!}{\includegraphics{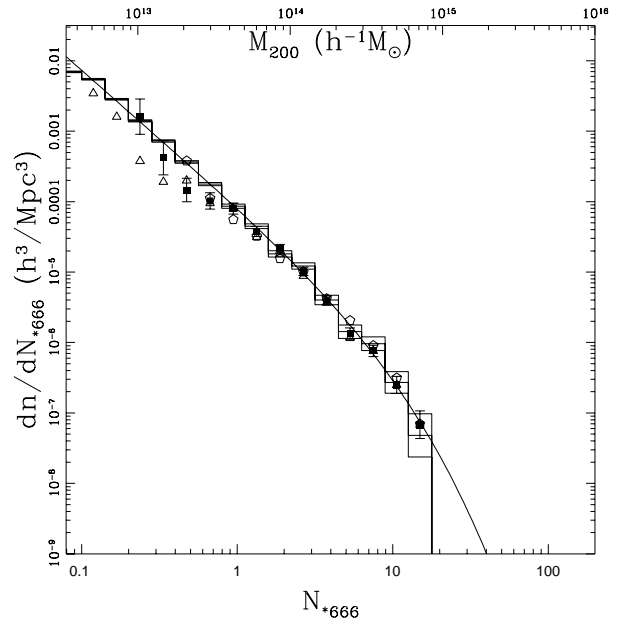}}
\end{center}
\caption{\footnotesize%
The number function $dn/dN_{*666}$ for the synthetic data.  The 
triangles, squares and pentagons show the results for $N_{\rm thresh}=3$,
$5$ and $10$ and their statistical errors. 
The solid histogram shows the actual distribution in the 
$(200h^{-1}{\rm Mpc})^3$ N-body cube and its Poisson errors for the
same bins. The smooth solid curve shows the expected distribution 
based on the Jenkins et al.~(\protect\cite{JFWCCEY}) CDM mass 
function and the synthetic catalog's multiplicity function. 
The mass scale at the top is based on the true occupation 
function for the synthetic data.
}
\label{fig:numfunc}
\end{figure}

\begin{figure}
\begin{center}
\resizebox{3.3in}{!}{\includegraphics{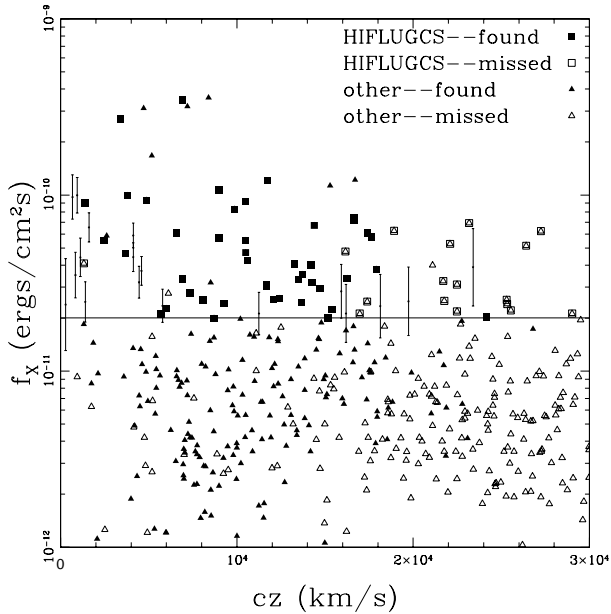}}
\end{center}
\caption{\footnotesize%
Detection of X-ray clusters as a function of redshift and X-ray flux $f_X$.
The solid points are X-ray clusters that were found and the open points
are X-ray clusters which were missed.  The squares show the clusters from
the HIFLUGCS survey which is complete to $f_X=2\times10^{-11}$~ergs/cm$^2$~s
(the horizontal line). The triangles show all clusters with X-ray
flux measurements that we matched to our output catalog.  The missed
HIFLUGCS target at low redshift is the NGC~4636 group, and it is probably
a failure in our matching methods rather than a true non-detection.
}
\label{fig:hiflugcs}
\end{figure}

\subsection{The Cluster Number Function \label{sec:numfunc}}

Finally we examine how well we can recover the cluster number function
$dn/dN_{*666}$ of the synthetic catalog.
Fig.~\ref{fig:numfunc} shows the distributions estimated using thresholds
of $N_{\rm thresh}=3$, $5$ and $10$ in defining $V_{\rm max}$
(see \S\ref{sec:dndN}).
The three estimates are mutually consistent, extending over two decades in
$N_{*666}$.
We can also compute the distribution for the $(200h^{-1}{\rm Mpc})^3$ N-body
simulation volume from which our synthetic catalog is drawn.
The characteristic volume for fair samples of the universe is
roughly $V_0=(100h^{-1}{\rm Mpc})^3$, which we shall use as a unit of volume.
In these units the simulation has a volume of $8V_0$, and to generate the
synthetic catalogs we periodically replicate the simulation.
The number function of low $N_{*666}$ clusters is derived from a small region
of the simulation, while the number function of high $N_{*666}$ clusters 
is drawn from several of the periodic repetitions of the data.
There is also good agreement with the semi-analytic number function found by
combining the the synthetic catalog's multiplicity function with the
Jenkins et al.~(\cite{JFWCCEY}) CDM mass function following the procedures in
\S\ref{sec:dndN} (Eq.~\ref{eqn:dndNanal}).
The only significant difference is the absence of high $N_{*666}$ clusters,
but this is a real feature of the N-body simulation used to produce the
synthetic catalog.  If we combine the mass function and the number function
to estimate the occupancy function, we obtain mutually consistent estimates
of $N_{*666}(M_{200})$ for $N_{\rm thresh}=3$, $5$ and $10$ which have
slopes and normalizations slightly above (1--2$\sigma$) the true relation
(see Table~\ref{tab:multiplicity}).

\section{2MASS Clusters} \label{sec:clusters}

With this understanding of the algorithm, we can now search for
clusters in the real 2MASS data.  We present no catalog at present
pending a reanalysis of the full 2MASS data to a fainter limiting
magnitude than is presently available.  Instead, we focus on the 
properties of the clusters.  In any fitted relation, we have selected 
the cluster sample used to perform the fit in exactly the same manner 
as was used for the synthetic catalog, and the order of the analyses
parallels that used to test the algorithm on the synthetic catalog.
Our standard sample consists of all galaxies identified from the
2MASS extended source catalog (see Jarrett et al.~\cite{Jarrett00})  
with $|b|> 5^\circ$ to an extinction corrected magnitude limit
of K$\leq 12.25$~mag.

\subsection{Tests of Completeness}

We start with two tests of the completeness of the algorithm for 
the real data.  The first test compares the ``shallow" cluster catalogs 
generated using a bright subsample of the galaxy catalog,
truncated at K$\leq 11.25$~mag, to the ``deep" cluster catalogs generated
using the complete sample to K$\leq 12.25$~mag.  
The second test is to compare our cluster catalog to local
X-ray surveys.

\subsubsection{Comparison of shallow and deep surveys}

We first discuss the results from comparing the shallow and deep
cluster catalogs.  With a one magnitude change in the survey depth,
individual clusters will have significant changes in their 
membership, as a cluster which
contained only $L>L_*$ ($L>0.1L_*$) galaxies in the bright
subsample, would on average contain 3.4 (1.5) times as many
galaxies in the complete sample.  
Using the same termination
criterion for the search, the shallow catalog contains only
375 clusters, as compared to 1743 in the deep catalog.  

We first consider the reliability of the catalog by trying to
identify shallow catalog clusters which have no counterpart
in the deeper catalog.   By cross-referencing the cluster assignments 
of the galaxies in the two catalogs we estimated a matching 
probability for identifying a cluster in the shallow catalog with 
each cluster in the deep catalog.  We included all galaxies 
with a 50\% or higher cluster membership probability.
The cluster matching probability can range from 0, if
no galaxy assigned to a shallow cluster has been assigned
to a deep cluster, to unity if every galaxy assigned to the
shallow cluster is assigned to the same deep cluster.  
Only 4 of the 375 clusters lack a firm counterpart in
the deeper catalog (defined by a matching probability
$<50\%$), and an additional 8 may have ambiguous
assignments (matching probability of 50--70\%). Most
shallow clusters have unambiguous counterparts, with
78\%, 85\% and 91\% of the shallow clusters having
match probabilities exceeding 95\%, 90\% and 85\%
respectively.  The match probabilities are not unity
because galaxies with modest membership probabilities
can be dropped or assigned to a different cluster 
based on the deeper catalogs, reducing the match
probability, even if the galaxies in the cluster 
cores remain perfectly matched.  Thus,   if we
define a false positive as a cluster which will lack
a clear counter part in a deeper survey, our method
appears to have a false positive rate of 1--3\% 
depending on the desired threshold. Be warned, however,
that this false positive rate is not the same as the
true false positive rates we considered for the synthetic
data -- a real but unvirialized overdensity of galaxies can 
be identified as a false positive in the synthetic data, but 
our deeper survey may still identify the clump as a cluster.

Next we examine the relative completeness of the two catalogs.
We divided the deep catalog into bins of $N_{*666}$ and 
computed the fraction of the deep catalog clusters found by
the shallow survey as a function of redshift.  For each
richness bin, the shallow survey is at least 80\% 
complete until the redshift at which the least rich
clusters in the bin are expected to have fewer than
3 galaxies.  For example, for the clusters in the 
deep catalog with $1 < N_{*666} < 3$, corresponding
to masses near $10^{14}h^{-1}M_\odot$, the completeness
of the shallow catalog is 92\% (61\%) for redshifts
below that where an $N_{*666}=1$ ($3$) cluster is 
expected to contain 3 galaxies.   The deep catalog
presumably has similar properties.

The final comparison we make is to compare the estimates of 
$N_{*666}$ for the 363 well-matched clusters from the two 
catalogs.  We find that 
\begin{equation}
  \log N_{*666}^{\rm shallow} =
  (0.07 \pm 0.01) + (1.00\pm0.01) \log N_{*666}^{\rm deep}
\end{equation}
with a variance of about the relation of $0.21$ dex. Formally,
the error estimates for $N_{*666}$ would have to be raised by
14\% to make the variance consistent with the error estimates.
The shallow catalog estimates of $N_{*666}$ are higher by 17\% 
on average (also found as the median difference).  The offset
is probably a real systematic effect -- the shallow catalog
will preferentially include clusters with excess numbers of
$K<11.25$~mag galaxies because an upward Poisson fluctuation
increases the likelihood of including the cluster in the
catalog.  The same cluster is unlikely to show the same
upward fluctuation in the $K<12.25$~mag catalog, producing
a net bias.  This interpretation predicts that a comparison
restricted to clusters in the shallow survey with more
galaxies, corresponding to clusters with smaller uncertainties
in $N_{*666}$, should have less of a bias.  For example, if
we restrict the comparison to the 41 clusters in the shallow 
catalog with errors in $\log N_{*666}$ smaller than
$0.10$, the average offset is reduced to 7\%
($0.03\pm0.02$) from 17\%.  

\subsubsection{Comparison with X-ray catalogs}

The second test compares our complete catalog to X-ray selected
samples.  It is difficult to do so in a completely satisfactory
manner because the selection methods are so different.  We 
first compare our results to the 63 clusters in the HIFLUGCS 
X-ray flux limited sample ($f_X > 2\times 10^{-11}$~ergs/cm$^2$~s, 
Reiprich \& B\"ohringer~\cite{Reiprich02}).  The HIFLUGCS survey 
covers most of the sky with $|b|>20^\circ$ plus small holes
near the LMC, the SMC and the Virgo cluster.  Fig.~\ref{fig:hiflugcs}
shows that we find all HIFLUGCS clusters out to the redshifts where 
we expect even rich clusters to have too few galaxies to be 
detected by our algorithm ($cz \simeq 15000$~km/s).  Our
automated matching system identified counterparts to 97\%
(35 of 36) of the HIFLUGCS clusters with $cz < 15000$~km/s.
We missed one very low redshift system, the NGC~4636 
group at $cz=1320$~km/s, but closer examination strongly 
suggests that this is a matching and coordinate problem
rather than a genuine failure.  Between 15000~km/s and 20000~km/s
we find 67\% (8 of 12) of the HIFLUGCS clusters.   
We miss the lower temperature clusters ($\langle T_X \rangle =4.3$~keV)
and find the higher temperature clusters ($\langle T_X \rangle =6.6$~keV).
At higher redshifts, an increasing fraction of X-ray clusters are
genuinely missed. Spot checks of these regions show only 0--2 
galaxies at the position and redshift of the X-ray cluster.

We can also use our $N_{*666}$ estimates to estimate an X-ray flux
(based on the correlation between $N_{*666}$ and the X-ray luminosity
in Eq.~\ref{eqn:NLx}).  This leads to 18 (14 below $cz=15000$~km/s) clusters 
whose $N_{*666}$ values and coordinates suggest the cluster should 
have an X-ray flux large enough to be included in HIFLUGCS.  These
flux estimates are not very accurate, but we show their distribution,
in Fig.~\ref{fig:hiflugcs}.  They can be divided into three
loose categories.  First, there are 7 very low redshift systems
with $cz <  2000$~km/s.  Some have very high likelihoods in our
catalog (the two most prominent are the Eridanus group and the 
Ursa Major cluster which are the 26th and 38th clusters found),
but no published X-ray studies.  These systems have virial sizes
significantly larger than the ROSAT field of view.  Second, there
are 5 at intermediate redshifts $cz \sim 5000$~km/s.  The two
most prominent systems here, Abell S0805 and Abell 3574,
are the 9th and 17th clusters found.  Finally, there are 6 systems 
above $cz >10000$~km/s of which the most prominent are Abell 1913 and 
Abell S0740.  After considerable and time consuming effort, we gave
up on tracking down the origins of these objects.  Many of the 
systems are well studied clusters with X-ray emission (e.g.~Abell S0805)
that do not appear in any of the ROSAT catalogs we have used for our comparisons.
Some are systems contaminated by X-ray emission from AGN (e.g.~Abell 3574).  The very low
redshift systems have angular extents comparable (or larger) than
the ROSAT aperture. In short, understanding these systems in detail requires
a careful analysis of the ROSAT data which is beyond the scope of
our present effort.  We note, however, that Donahue et al.~(\cite{Donahue02})
also find examples of optically detected clusters which appear to
have fainter than expected X-ray counterparts.

Fig.~\ref{fig:hiflugcs} also shows all other clusters for which we
have found a ROSAT X-ray flux measurement but are not derived from
complete, flux-limited samples.  As these are largely
derived from the RASS survey as well, they tend to have X-ray
fluxes $f_X \gtorder 2\times 10^{-12}$~ergs/cm$^2$~s.  The 
automatic matches find 84\% (178 of 213) of the clusters with X-ray 
flux measurements in our input local cluster catalog and $cz < 15000$~km/s. 
As with the HIFLUGCS sample, some of the mismatches at low redshift are due 
to cross-matching and coordinate problems.  

In summary, by comparing our standard cluster catalog to either a
shallow catalog or X-ray cluster catalogs, we can show that we 
achieve very high completeness up to the redshifts where clusters
contain too few galaxies for robust detection (about 3 galaxies).
It is difficult, however, to use either comparison to robustly
estimate a false positive rate.  The comparison of the shallow
and deep catalogs can only set a lower bound (of a few percent)
on the false positive rate.  The HIFLUGCS sample sets an upper
bound of 40\%, but this mainly reflects the limitations of the
RASS and our ability to estimate X-ray fluxes from $N_{*666}$
rather than a true estimate of the false positive rate. 

\begin{figure}
\begin{center}
\resizebox{3.3in}{!}{\includegraphics{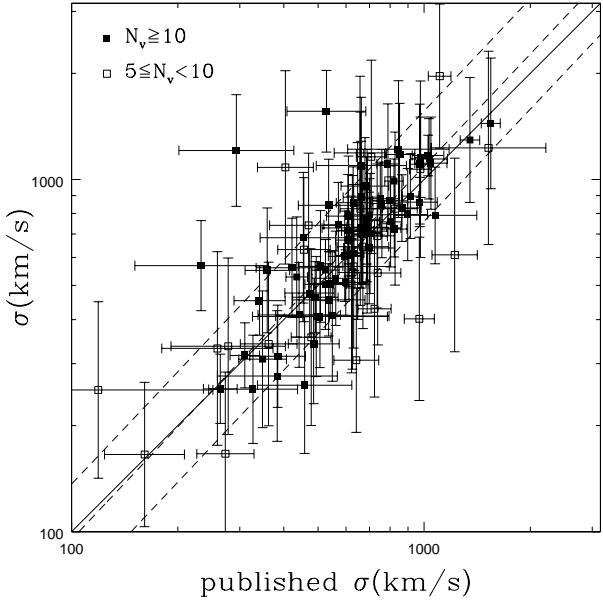}}
\end{center}
\caption{\footnotesize%
Comparisons between our velocity dispersion estimates and published
estimates.  Filled points have $N_v \geq 10$ velocities and
open points have $5 \leq N_v < 10$ velocities. The dashed lines
show the best fit relation and its width, which would lie on
the solid line if the agreement was perfect.  }
\label{fig:sigsig}
\end{figure}

\begin{figure}
\begin{center}
\resizebox{3.3in}{!}{\includegraphics{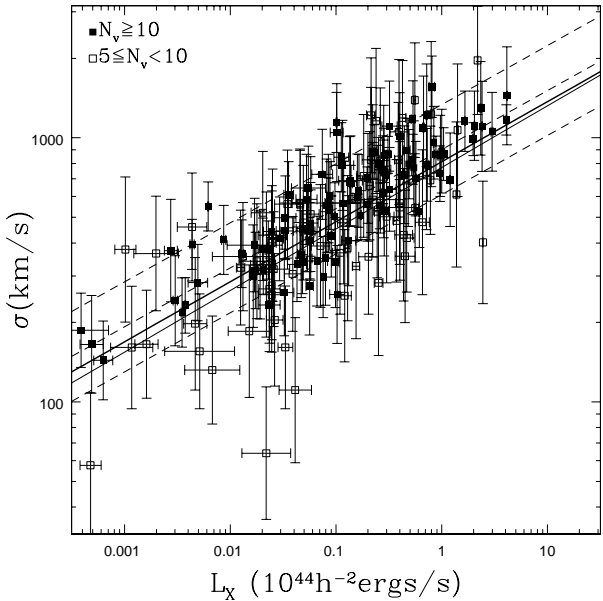}}
\end{center}
\caption{\footnotesize%
The velocity dispersion estimate $\sigma$ as a function of the X-ray
luminosity $L_X$. Filled points have $N_v \geq 10$ velocities and
open points have $5 \leq N_v < 10$ velocities.  The light solid line is
the correlation from Mulchaey \& Zabludoff~(\protect\cite{Mulchaey98})
and the heavy solid line is the correlation from
Mahdavi \& Geller~(\protect\cite{Mahdavi01}).}
\label{fig:sigxlum}
\end{figure}

\begin{figure}
\begin{center}
\resizebox{3.3in}{!}{\includegraphics{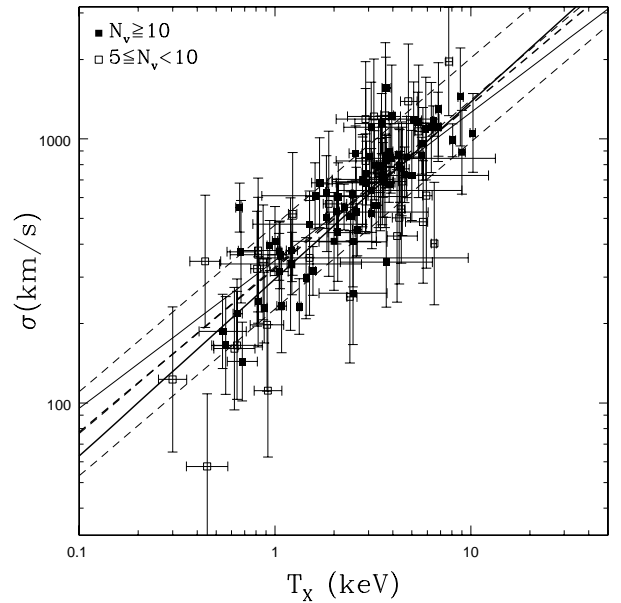}}
\end{center}
\caption{\footnotesize%
The velocity dispersion estimate $\sigma$ as a function of the X-ray
temperature $T_X$. Filled points have $N_v \geq 10$ velocities and
open points have $5\leq N_v < 10$ velocities.
The light solid line is the fit from Wu, Fang \& Xu~(\protect\cite{WuFanXu})
with slope 0.56, the heavy solid line the fit with slope 0.67. The
heavy dashed line is the estimate from Girardi et al.~(\protect\cite{Gir98}).
}
\label{fig:sigxtemp}
\end{figure}

\subsection{Further Checks of the Velocity Dispersion}

Before proceeding to estimates of the halo occupancy function, we made
three additional checks of our velocity dispersion estimates, comparing
directly with other published values and through comparisons to X-ray
observations of the clusters.  All fits of variable correlations are
performed as $\chi^2$ fits including the errors in both variables and
including all clusters with at least $N_v\geq 5$ associated redshifts
that we could match to the other data.

First we checked our velocity dispersions against the velocity dispersions
$\sigma_{\rm pub}$ from Girardi et al.~(\cite{Gir98}; Table 2) and 
Wu et al.~(\cite{WuXuFa}).  We matched 98 clusters 
with at least $N_v\geq 5$ redshift measurements to these previous 
estimates, and we found good agreement within the 
rather large errors (Fig.~\ref{fig:sigsig}).  If we fit
a power-law, our $\sigma$ is related to $\sigma_{\rm pub}$ by
\begin{equation}
  {\sigma \over 1000\hbox{km/s} } 
 = (1.09\pm0.05)
  \left({\sigma_{\rm pub}\over 1000\hbox{km/s}}\right)^{1.06\pm0.08}
\end{equation}
with a dispersion of $0.16$~dex that is consistent with the
uncertainties. If we restrict the fit to the 69 systems with
$N_v\geq 10$ redshift measurements, we find essentially the
same parameters.  Hence our membership probability weighting method
for determining velocity dispersions is consistent with the more
standard methods based on rejecting outliers.

We also find 173 (108) clusters with X-ray luminosities (temperatures)
and at least $N_v\geq 5$ galaxies with redshifts from
from the sources summarized in \S\ref{sec:MassEst}.  
Figs.~\ref{fig:sigxlum} and \ref{fig:sigxtemp} show our estimates
of the relations between the X-ray luminosity $L_X$ and the X-ray 
gas temperature $T_X$ to the velocity dispersion estimates.  
The correlation between the velocity dispersion and the X-ray
luminosity, including errors in both quantities, is 
\begin{equation}
    \log L_{44}   = (0.20 \pm 0.07) + (4.47\pm0.24) 
      \log\left( { \sigma \over 10^3\hbox{km/s} }\right)
    \label{eqn:sigmaL}
\end{equation}
where $L_{44}=L_X h^2/10^{44}\hbox{ergs/s}$.
The variance in the X-ray luminosity at fixed velocity dispersion
is a (typically) large 0.76~dex.  If we restrict the fit to systems
with $N_v\geq 10$ the normalization of $0.18\pm0.08$ is little
changed, but the slope
steepens to $4.69\pm0.28$ with a reduced variance of 0.55~dex.  
The slope and normalization are
close to those found by Mulchaey \& Zabludoff~(\cite{Mulchaey98},
zero point $0.48\pm 1.09$, slope $4.29\pm0.37$) or
Mahdavi \& Geller~(\cite{Mahdavi01}, zero point $0.40^{+0.9}_{-2.0}$, slope
$4.4^{+0.7}_{-0.3}$).  Wu et al.~(\cite{WuXuFa}) found a range of slopes
from $2.56\pm0.21$ to $5.24\pm0.29$ depending on their fitting methodology.
The zero-point uncertainties we cite for these published relations are 
overestimates because we had to correct the relations to a velocity
scale $10^3$~km/s from one of $1$~km/s without knowing the full
covariance matrix of the fit.  For the 108 clusters with 
X-ray temperatures, we find
\begin{equation}
    \log\left( { \sigma \over 10^3\hbox{km/s} }\right) =
      (0.01 \pm 0.02) +
      (0.63\pm0.04) \log\left( { T_X \over 6\hbox{keV} }\right)
    \label{eqn:sigmaT}
\end{equation}
with a spread of 0.16~dex, which is close to the relations found by
Mulchaey \& Zabludoff~(\cite{Mulchaey98}, zero point $-0.01\pm 0.04$,
slope $0.51\pm0.05$),
Wu et al.~(\cite{WuXuFa}, zero point $-0.00\pm 0.03$, slope $0.65\pm0.02$),
and Girardi et al.~(\cite{Gir98}, zero point $-0.01\pm 0.03$, slope
$0.62\pm0.04$).
Fitting to the 79 systems with $N_v\geq 10$ gives the same parameters
($0.01\pm0.02$ and $0.62\pm0.04$) with a reduced scatter of 0.12~dex.
The similarity with these published analyses suggests that our method for
determining velocity dispersions is comparable to those used for other studies
despite the small numbers of galaxies used in our dispersion estimates compared
to these more extensive surveys of individual clusters.

\subsection{The Halo Multiplicity Function} \label{sec:multiplicity}

\begin{figure}
\begin{center}
\resizebox{3.3in}{!}{\includegraphics{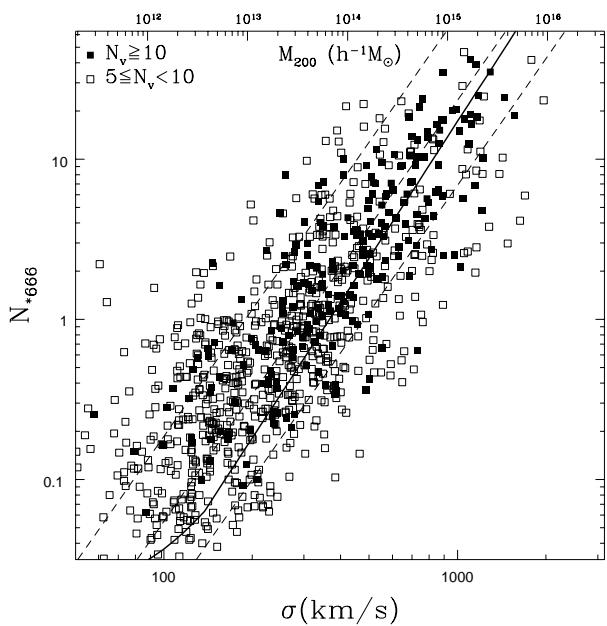}}
\end{center}
\caption{\footnotesize%
The number of galaxies $N_{*666}$ as a function of the cluster velocity
dispersion $\sigma$ for the 2MASS data.
Points and lines as in Fig.~\protect\ref{fig:numsig}.}
\label{fig:numsig2mass}
\end{figure}

\begin{figure}
\begin{center}
\resizebox{3.3in}{!}{\includegraphics{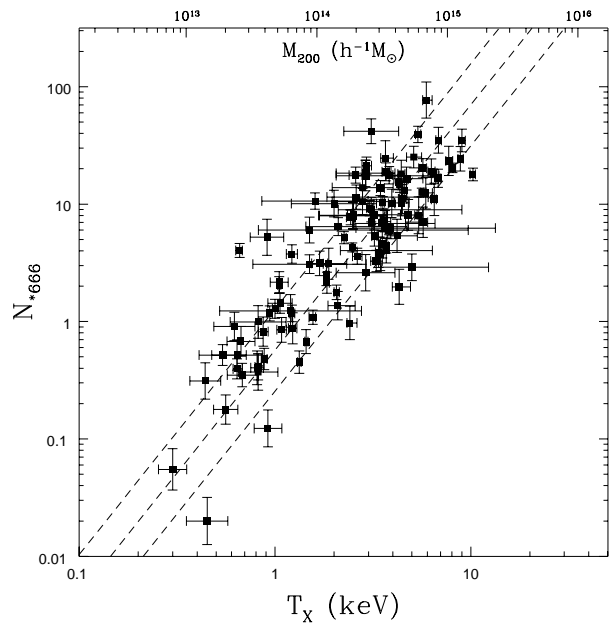}}
\end{center}
\caption{\footnotesize%
The number of galaxies $N_*$ versus the X-ray temperature $T_X$. The 
dashed lines show the best power law fit for $T_X\geq 1$~keV and the 
average dispersion of the data about the fit. }
\label{fig:numxtemp}
\end{figure}

We now derive the halo multiplicity function based on the three observables,
velocity dispersion, X-ray luminosity and X-ray temperature, which we can
later relate to the cluster mass.  We first derive the observed
correlations and then estimate the conversion to the desired correlation
with cluster mass.   

Fig.~\ref{fig:numsig2mass} shows that the observed relationship between
velocity dispersion and galaxy number is very similar to that in the
synthetic data (\S\ref{sec:tests}, Fig.~\ref{fig:numsig}).  The best
fit relation for the 939 systems with $N_v\geq 5$, including the errors 
in both quantities, is
\begin{equation}
  \log N_{*666} = (1.37\pm 0.03) + (2.63 \pm 0.06)\log(\sigma/10^3\hbox{km/s})
  \label{eqn:Nsigma}
\end{equation}
compared to $1.32\pm0.04$ and $2.76\pm0.08$ for the zero point and slope of the
fits to the synthetic catalog ($1.41$ and $2.75$ for the input relation).
Only one system is dropped from the fits.
The scatter in $N_{*666}$ at fixed dispersion
is slightly larger in the real data ($0.53$~dex versus $0.50$~dex). 
If we restrict the fit to the 238 systems with $N_v \geq 10$, we obtain
consistent parameters ($1.31\pm0.04$ and $2.60\pm0.10$) with a reduced 
scatter of $0.41$~dex. 
While the significance of the fit in Eq.~(\ref{eqn:Nsigma}) is quite high,
the very large scatter makes it hard to see by eye.  Therefore as a check we
also used the dispersion in the distribution of galaxy-group velocities in
`stacked' clusters, binned in 5 logarithmic bins of $N_{*666}$, to make the 
trend more clearly visible.
We obtained a fit very close to Eq.~(\ref{eqn:Nsigma}).

\begin{figure}
\begin{center}
\resizebox{3.3in}{!}{\includegraphics{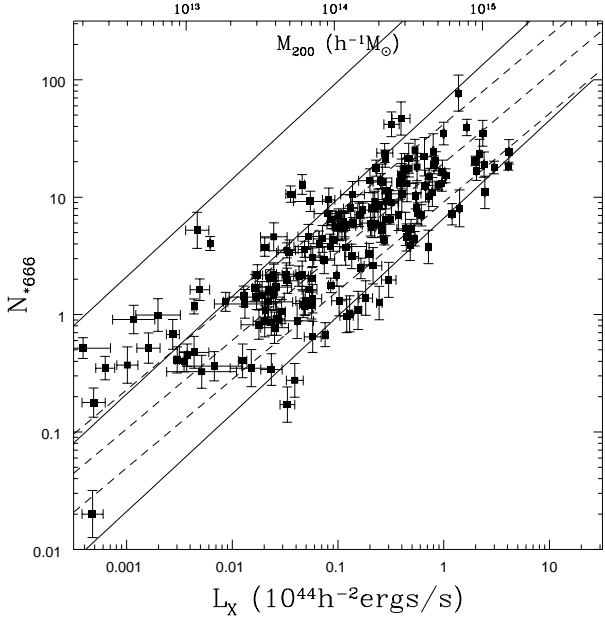}}
\end{center}
\caption{\footnotesize%
The number of galaxies $N_*$ versus the X-ray luminosity $L_X$.
The dashed curves show our best power-law fit to the relation
for $L_X \geq 10^{42}h^2$~ergs/s.  The solid curves show our 
theoretically estimated conversion (Eq.~{\protect\ref{eqn:postman2}}) 
of the $\Lambda_{cl}$--L$_X$ correlation found by Donahue et 
al.~(\protect\cite{Donahue01}) and the uncertainties in its 
normalization.  We can match the two relations by raising the
normalization of the conversion to $\Lambda_{cl}\simeq 21 N_{*666}^{3/4}$. }
\label{fig:numxlum}
\end{figure}

Next we compare the halo occupation number to the X-ray properties of the
clusters, as shown in Figs.~\ref{fig:numxtemp} and \ref{fig:numxlum}.
We must face two problems in correlating $N_{*666}$ with the X-ray 
data.  First, we find that the scatter in the correlations is 
significantly larger than is consistent with the uncertainties in
either quantity.  In part this is due to the presence of significant
intrinsic scatter in the X-ray properties of clusters.
Second, there is considerable evidence for a break in the X-ray 
properties between groups and clusters, generally believed to
be due to significant cooling and heating from star formation processes
in the lower mass systems
(David et al. \cite{David93};
 Ponman et al. \cite{Ponman96};
 White, Jones \& Forman \cite{WJF97};
 Allen \& Fabian \cite{AF98};
 Markevitch \cite{Mar98};
 Arnaud \& Evrard \cite{ArnEvr};
 Helsdon \& Ponman \cite{Helsdon00};
 Finoguenov, Reiprich \& B\"ohringer \cite{FinReiBoh}).
Between the intrinsic scatter and the break in the properties, the parameters
for power-law fits to X-ray correlations can depend on the fitting methods
and the data used.

In order to include the effects of these problems we made two modifications to
our fitting procedures. First, we added a logarithmic systematic error
(or intrinsic scatter) $\sigma_{\rm sys}$ in quadrature with the measurement
errors for the fitted variables.
We then adjusted $\sigma_{\rm sys}$ so that the fit has $\chi^2=N_{\rm dof}$.
Second, we fit power-law relations including the X-ray data only for the more
massive clusters with $T_X \geq 1$~keV and $L_X \geq 10^{42}h^2$~ergs/s.
This reduces the biases in the correlations from attempting to fit the lower
mass systems where non-adiabatic effects begin to modify the X-ray properties
while leaving enough systems to obtain a statistically significant fit.
As in all our standard fits we include only the systems with 
$N_v \geq 5$ associated redshifts.


If we now find the best power-law relation between $L_X$ and $T_X$, we find
\begin{equation}
   \log L_{44} = (0.21\pm0.04)
      + (2.66\pm0.12) \log \left( { T\over 6\hbox{keV}}\right)
  \label{eqn:LxTx2}
\end{equation}
based on 84 clusters with $\sigma_{\rm sys}=0.08$ and a mean scatter of
$0.30$~dex.  The normalization is consistent with our adopted standard
from Markevitch~(\cite{Mar98}) in Eq.~(\ref{eqn:LxTx}), but it has
a significantly steeper slope.  If we include no limits on the
X-ray properties we find a similar zero-point of $0.28\pm0.05$ 
but a still steeper slope of $3.04\pm0.10$ and a larger scatter of $0.35$~dex.
This fit matches that of Xue \& Wu~(\cite{Xue00}) under similar assumptions.
There is clearly a break in the slope of the $L_X$--$T_X$ relation
 near $T_X=1$~keV which is incompatible
with a single power law providing a good fit.

We first fit the number-temperature relation as a power law using 84
clusters, as shown in Fig.~\ref{fig:numxtemp}, to find that
\begin{equation}
 \log N_{*666} = (1.38\pm 0.05) + (2.09 \pm 0.17) \log (T/6\hbox{keV}).
  \label{eqn:NTx}
\end{equation}
with $\sigma_{\rm sys}=0.12$~dex and a mean scatter of $0.35$~dex.  This
relationship changes little if we drop the restrictions on the X-ray
properties.  The fit for all 102 clusters with X-ray temperatures, 
luminosities and $N_v\geq5$, has a zero-point of $1.39\pm0.05$, a
slope of $1.97\pm0.12$, and a negligible increase in $\sigma_{\rm sys}$
and the scatter.

Fig.~\ref{fig:numxlum} shows the results for fitting the 153 clusters with
$L_X\geq 10^{42}h^{-2}$~ergs/s and $N_v \geq 5$. We find
\begin{equation}
 \log N_{*666} = (1.29\pm 0.05) + 
                 (0.75\pm 0.05)\log L_{44}.
  \label{eqn:NLx}
\end{equation}
for $\sigma_{\rm sys}=0.24$ and with a scatter of $0.33$~dex.  If we
include all 173 clusters with X-ray luminosity measurements and
$N_v \geq 5$, the zero-point becomes $1.23\pm0.04$ and the slope 
of $0.64\pm0.03$ is significantly flatter.  The distribution of
clusters in Fig.~\ref{fig:numxlum} is clearly inconsistent with
a simple power law if we extend the fit to low X-ray luminosities,
so we will not consider this case further.  
As a consistency 
check, we can combine the $N_{*666}$--$T_X$ and $N_{*666}$--$L_X$
relations (Eqs.~\ref{eqn:NLx} and \ref{eqn:NTx}) to infer 
a zero-point of $0.12\pm 0.09$ and a slope of $2.79\pm0.29$ 
for the $L_X$--$T_X$ relation which are consistent with the
direct fit in Eq.~(\ref{eqn:LxTx2}).  Similarly, if we combine
the $N_{*666}$--$\sigma$ and $N_{*666}$--$T_X$ relations
(Eqs.~\ref{eqn:Nsigma} and \ref{eqn:NTx}), we infer a zero-point
of $0.00\pm0.02$ and a slope of $0.79\pm0.07$ for the
$\sigma$--$T_X$ relation which are reasonably consistent with 
the direct fit in Eq.~(\ref{eqn:sigmaT}).   

We can also compare our results to those of Donahue et al.~(\cite{Donahue01}),
who compared the properties of clusters found in X-ray surveys with those
found using the Postman et al.~(\cite{MF}) matched filter algorithm.
In this algorithm, clusters are characterized by their total luminosity
in units of $L_*$, $\Lambda_{cl}$, where we estimated a conversion of
$\Lambda_{cl}=11N_{*666}^{0.75}$ (see Eq.~\ref{eqn:postman2}) between
our cluster normalizations.  Given this relation, we can convert their
estimated correlation with X-ray luminosity into our units, 
\begin{equation}
 \log N_{*666} = (1.8\pm 1.0) +
                 (0.8\pm 0.2) \log L_{44}.
\end{equation}
While the slope and the normalization are consistent with our estimate
given the uncertainties in their relation, the normalization is 
systematically higher.  If we estimate the conversion between
$\Lambda_{cl}$ and $N_{*666}$ by matching the X-ray correlations
for the two variables, we find that $\Lambda_{cl}=21 N_{*666}^{0.83}$,
where the slope is consistent with our theoretical estimate 
(Eq.~\ref{eqn:postman2}) but the normalization is nearly doubled.  
Since the sense of the change is the same as that needed to better
match $\Lambda_{cl}$ and $N_{*666}$ for clusters of the same Abell
richness class (see \S3.6), we will adopt an empirical relation 
for the conversion of 
\begin{equation}
  \Lambda_{cl} \simeq 21 N_{*666}^{3/4}
   \label{eqn:postman3}
\end{equation}
for subsequent comparisons.  Using the original $L_X$ and $\Lambda_{cl}$
data points, we confirmed that this empirical conversion was correct,
while the theoretical conversion from Eq.~(\ref{eqn:postman2}) showed
an offset whose origin we do not fully understand.

The final step in the analysis is to convert the observed occupancy
functions ($N_{*666}(\sigma)$, $N_{*666}(T_X)$ and $N_{*666}(L_X)$) into 
an estimate of the occupancy function $N_{*666}(M_{200})$ as a function 
of the virial mass.  The results, summarized in Table~\ref{tab:multiplicity},
are dominated by systematic errors associated with the choice of the
mass scale.  The average normalization $A=1.25\pm0.03$ and slope
$B=1.02\pm0.10$ found when we set the mass scale using the 
velocity dispersion ($\sigma(M_{200})$, Eq.~\ref{eqn:Msigma})
are lower than the average normalization $A=1.58\pm0.02$ and
slope $B=1.21\pm0.14$ found when we set the mass scale using
the X-ray temperatures ($M_{200}(T_X)$, Eq.~\ref{eqn:MTx}).
The results based on the same mass scale are largely independent of the
observed occupancy function used to make the estimate.
This is particularly true of the normalization $A$, where the scatter between
the individual estimates is less than the formal errors, and less true of the
slope $B$ where the scatter between the individual estimates is somewhat
larger than the formal errors.
A shallower mass-temperature relation with $T\sim M^{1/2}$ rather than
$T\sim M^{2/3}$, perhaps better suited to the mix of high and low temperature
clusters we use, would bring the two estimates of the slopes into agreement.
In addition to these systematic differences, recall that in the synthetic data
our fits to the occupancy function underestimated the normalization by
$(19\pm7)\%$ and the slope by $0.06\pm0.03$
(see \S\ref{sec:hof} and Table~\ref{tab:multiplicity}).

\subsection{The Cluster Number Function}

\begin{figure}
\begin{center}
\resizebox{3.3in}{!}{\includegraphics{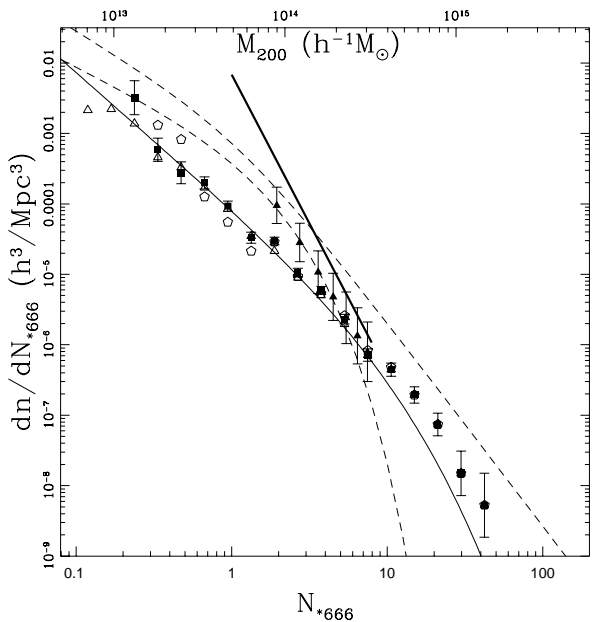}}
\end{center}
\caption{\footnotesize%
The number function $dn/dN_{*666}$ for the 2MASS sample.
The open triangles, filled squares and open pentagons show the results
for $N_{\rm thresh}=3$, $5$ and $10$.  In many cases these points overlap.
The bootstrap error estimates are shown only for $N_{\rm thresh}=5$.
The smooth curve is the number function expected for the synthetic
catalog in Fig.~\protect\ref{fig:numfunc}.  The heavy solid line and the
filled triangles show the $dn/d\Lambda_{cl}$ number functions from 
Donahue et al.~(\protect{\cite{Donahue01}}) and 
Postman et al.~(\protect{\cite{Postman02}}) respectively, converted to our
units using the empirical conversion of Eq.~({\protect\ref{eqn:postman3}}).
The lower dashed curve cutting off sharply at high mass shows the luminosity
function from the NOG in Marinoni et al.~(\protect\cite{Marinoni01}) converted
to a number function as described in the text.  The other dashed curve shows
the fit to the CfA survey luminosity function from
Moore, Frenk \& White (\protect\cite{MooFreWhi}) converted in the same way.
The equivalent mass scale shown at the
top of the figure is the conversion from $N_{*666}$ to $M_{200}$
estimated in Eq.~\protect\ref{eqn:finaloccfunc}. }
\label{fig:numfunc2}
\end{figure}

Next we estimate the cluster number function, $dn/dN_{*666}$, for the
2MASS sample as shown in Fig.~\ref{fig:numfunc2}. The number function
for the real sample extends to higher values of $N_{*666}$ than that
for the synthetic sample (see Fig.~\ref{fig:numfunc}) but otherwise
looks very similar.  The results for the three values of $N_{\rm thresh}$
are again mutually consistent.  Note that for large $N_{*666}$ the
number function is systematically above that of the synthetic catalog.

Comparing our results with previous estimates of the same quantity is
quite difficult, due to different conventions for estimating `richness'.
The best we can do is to apply reasonable transformations based on scaling
relations between the widely differing definitions, realizing that this
method is far from perfect.

Donahue et al.~(\cite{Donahue01}) and Postman et al.~(\cite{Postman02})
have estimated the density of clusters at intermediate redshift in
terms of $\Lambda_{cl}$ (see \S\ref{sec:MassEst} and Eq.~\ref{eqn:postman})
over the ranges $20 \ltorder \Lambda_{cl} \ltorder 100$ and
$\Lambda_{cl} \gtorder 30$ respectively.
They fit the number density as a power law of the form
\begin{equation}
  {dn\over d\Lambda_{cl}} = N_0 \left( {\Lambda_{cl}\over 40} \right)^{-\alpha}
\end{equation}
finding $N_0 = 6^{+3}_{-1} \times 10^{-6}h_{75}^3{\rm Mpc}^{-3}$ with
$\alpha=5.3\pm0.5$ and $N_0=(1.55\pm0.40)\times 10^{-6}h_{75}^3{\rm Mpc}^{-3}$
with $\alpha=4.40 \pm0.30 $ respectively.  If we use the empirical conversion
between $\Lambda_{cl}$ and $N_{*666}$ derived from comparing the two
richness estimates at fixed X-ray luminosity (Eq.~\ref{eqn:postman3}), then
we find good agreement for the number density of $N_{*666}\sim 10$  
clusters (see Fig.~\ref{fig:numfunc}).  If, however, we use our theoretical 
estimate of the 
conversion (see \S\ref{sec:MassEst}, Eq.~\ref{eqn:postman2}), then the
converted Donahue et al.~(\cite{Donahue01}) and
Postman et al.~(\cite{Postman02})
number densities are nearly an order of magnitude higher than our
own estimates.  Since both our catalogs (\S5.1) and 
Donahue et al.~(\cite{Donahue01}, \cite{Donahue02}) have similar
completeness compared to X-ray surveys, the need to use the empirical
conversion in order to obtain similar number densities suggests that
there is a problem with the theoretical conversion between the two
richness estimates.

Marinoni, Hudson \& Giuricin~(\cite{Marinoni01}) used the Nearby Optical
Galaxy (NOG, Marinoni \cite{NOG}) sample to make a similar estimate based
on the group catalogs of Giuricin et al.~(\cite{Giuricin00}).
If $\Lambda_G=L_{\rm tot}/L_*$ is the estimate for the total group luminosity
$L_{\rm tot}$ in units of the luminosity of an $L_*$ galaxy, they derive a
number function of
\begin{equation}
  { dn \over d\Lambda_G} = { n_0 \over \Lambda_{G0} }
   \left( { \Lambda_G \over \Lambda_{G0} } \right)^{-\alpha_G}
   \exp(- \Lambda_G/ \Lambda_{G0})
\end{equation}
where $\alpha_G \simeq -1.4$, $n_0= 4.8 h_{75}^3 \times 10^{-4}$~Mpc$^{-3}$,
and $\Lambda_{G0} \simeq 9.9$.  For a galaxy luminosity function with their
estimated Schechter slope of $\alpha=-1.1$, the equivalent number of $L>L_*$
galaxies is $N_{G0}\simeq \Lambda_{G0}/5.1 = 1.9$.
For groups selected with a standard FoF algorithm, we expect
$N_{*666}\simeq 0.66 N_G$, but the applicability of this conversion
to the Marinoni et al.~(\cite{Marinoni01}) estimates is just a rough
estimate.  The number function $dn/dN_{*666}$ is identical to the
Schechter function $dn/d\Lambda_G$ with $\Lambda_G$ replaced by $N_{*666}$ and 
$\Lambda_{G0}=1.3$.  This estimate, which is also shown in
Fig.~\ref{fig:numfunc2}, is a factor of 5 higher than ours for low $N_{*666}$
and cuts off more sharply.  The sharper cutoff is simply a consequence of
the $cz < 6000$~km/s redshift cutoff of the NOG Sample, which eliminates all
the nearby rich clusters.  The normalization could be brought into agreement
with our estimate if we have overestimated the coefficient in the 
relation $N_{*666}\simeq 0.66 N_G$ used to convert the FoF membership
into our standard overdensity.
Finally we have converted the luminosity function of the CfA sample, as
derived by Moore, Frenk \& White (\cite{MooFreWhi}), to $dn/dN_{*666}$ in
the same manner as described above and with the same difficulties.  
This is plotted as the other dashed
curve which agrees with the NOG estimate at small $N_{*666}$ but has
significantly more objects at high $N_{*666}$.  As with the comparison
to Donahue et al.~(\cite{Donahue01}) and Postman et al.~(\cite{Postman02}),
differences in the absolute number density are very sensitive to errors
in the conversion between richness estimates, and we lack any means of
doing this more precisely given the information available in 
Marinoni et al.~(\cite{Marinoni01}) and Moore et al. (\cite{MooFreWhi}). 

Table~\ref{tab:multiplicity} also gives the estimates for the halo occupancy
function needed to transform the Jenkins et al.~(\cite{JFWCCEY}) mass function
for our assumed cosmology into the observed number function (as in section
\ref{sec:numfunc}).  The results for $N_{\rm thresh}=3$, $5$ and $10$ are
mutually consistent and intermediate to the results from fitting the observed
occupancy functions in \S\ref{sec:multiplicity}.

\begin{figure}
\begin{center}
\resizebox{3.3in}{!}{\includegraphics{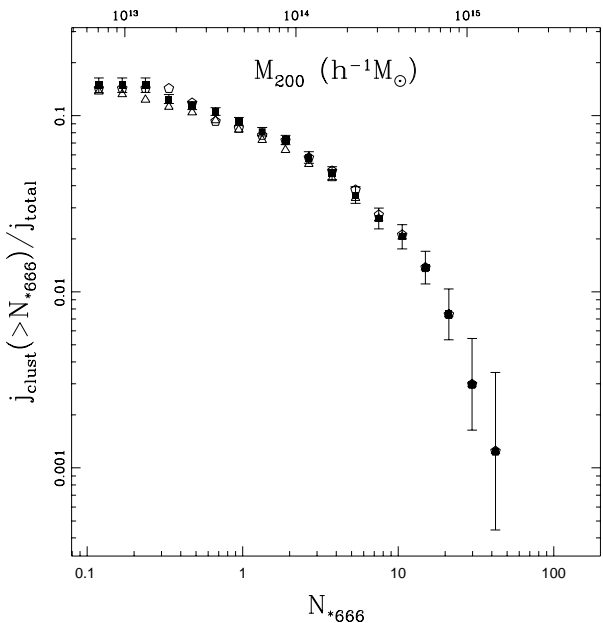}}
\end{center}
\caption{\footnotesize%
The fraction $j_{clust}(>N_{*666})/j_{total}$ of all stellar luminosity 
in systems with at least $N_{*666}$ galaxies.  The open triangles, 
filled squares and open pentagons show the results 
for $N_{\rm thresh}=3$, $5$ and $10$. In many cases these points
overlap. The
bootstrap error bars are shown only for $N_{\rm thresh}=5$.
The equivalent mass scale shown at the
top of the figure is the conversion from $N_{*666}$ to $M_{200}$ 
estimated in Eq.~\protect\ref{eqn:finaloccfunc}.}
\label{fig:lumfrac}
\end{figure}

\subsection{Derived Quantities}

Finally we discuss the implied infrared mass-to-light ratio of clusters
and the fraction of the local luminosity density associated with the
virialized regions of clusters.  We start by adopting a standard 
model for the halo occupation function.  Our uncertainties in the 
occupation function are dominated by systematic errors in how to
relate the observed quantities to cluster masses.  Depending on 
whether we set the mass scale using the velocity dispersion, the
X-ray temperature, or the cluster mass function, we find 
significant differences in both the normalization and the slope
of the occupancy function (see Table~\ref{tab:multiplicity}). 
We decided to simply average
the results for the three possible mass scales and use their 
dispersions as the standard errors:  
\begin{equation}
    \log N_{*666} = 
    (1.44\pm 0.17) + (1.10 \pm 0.09) \log (M_{200}h/10^{15}M_\odot).
    \label{eqn:finaloccfunc}
\end{equation}
The resulting errors are larger
than the statistical uncertainties for any given mass normalization
method and somewhat larger than the systematic offsets we found 
when estimating the occupancy function of the synthetic catalog.
The normalization is higher than we used in the synthetic catalog,
which seems reasonable given the visibly weaker fingers of god 
in the synthetic catalogs. 

From Eq.~(\ref{eqn:finaloccfunc}) we can estimate more traditional
quantities like the K-band mass-to-light ratios of the clusters.
Including the uncertainties in the luminosity function, the correction from
aperture to total magnitudes, and a zero point of $M_{\odot K_s}=3.39$~mag
(see Kochanek et al.~\cite{Kochanek01} and Cole et al.~\cite{Cole01})
we find that
\begin{equation}
   \left( { M_{200} \over L_{666} } \right)_K = (116 \pm 46)h 
      \left( { M_{200}h \over 10^{15}M_\odot } \right)^{0.10\pm0.09}
      \label{eqn:masstolight}
\end{equation}
where almost all the uncertainty comes from the uncertainty in the
normalization $A=1.44\pm0.17$ of the occupancy function (it would
be $(116\pm10)h$ for $A\equiv 1.44$).  This is nearly identical to
the baseline semi-analytic model of Cole et al.~(\cite{Cole00}) where 
the K-band mass-to-light ratio is $118(M_{200}h/10^{15}M_\odot)^{0.06}$
for clusters with masses above $10^{13.5}h^{-1}M_\odot$.  The mass-to-light
ratio is essentially independent of mass, which differs somewhat from
estimates in the B-band (e.g.~Girardi et al.~\cite{girardi00}, 
Marinoni \& Hudson~\cite{Marinoni01b}) which suggest a
steeper slope of $(M/L)_B \propto M^{0.1-0.3}$, albeit with 
significant uncertainties in the slopes.  If there is a real slope difference,
with $(M/L)_B/(M/L)_K \propto M^x$ then we can explain this as a trend in the
colors of the galaxy population with $\Delta (B-K) = 5 x \sim 0.5$
to $1.5$~mag over the range from $10^{13}M_\odot$ to $10^{15}M_\odot$.  All
general trends in cluster properties have the sense required to
produce the slope difference.  Starting from a constant K-band
mass-to-light ratio, the more massive clusters could be dimmer
in the B-band because of the increasing early-type galaxy
fraction (a B--K color shift of roughly 0.5~mag from Sa to E),
increasing metallicity ($\Delta($B--K$)/[Fe/H]\sim 1.2$~mag/dex) 
or increasing age ($\Delta($B--K$)/\log t \sim 0.5$ to $1.0$~mag/dex).
Like the halo occupation number itself, it is important to match definitions
when comparing our mass-to-light ratios to other estimates.
Our estimate is normalized to be the ratios of the two quantities in spheres of
radius $r_{N666}\simeq r_{M200}$, leading to higher mass to light ratios
than estimates comparing the light in cylinders to the mass in spheres.
For example, Rines et al.~(\cite{Rines01}) found $(M/L)_K=(75 \pm 23)h$ for
Coma using the 2MASS survey, and the Kochanek et al.~(\cite{Kochanek01})
luminosity function. At the $r_{N666}=1.5h^{-1}$~Mpc, which also matches the
Rines et al.~(\cite{Rines01}) estimate of the virial radius, we must
raise their estimate of the mass to light ratio by 25\% to 
$(M/L)_K=(92\pm 28)h$ to compare it to our estimates of the mass-to-light
in spheres.  This is quite close to our estimate of $(M/L)_K=(116\pm46)h$ 
for the mass-to-light ratio of $10^{15}h^{-1}M_\odot$ cluster like Coma
in Eq.~(\ref{eqn:masstolight}).

Since the K-band luminosity is well correlated with the total stellar
mass, we can use the cluster number function to obtain a rough estimate
of the fraction of stars in systems above a given mass or number threshold.
The total luminosity density of all galaxies is
$j_{\rm tot} = n_* L_* \Gamma[2+\alpha]$ while the equivalent luminosity
density contained clusters with $N_{*666} > N$
(and in the region with $r < r_{N666}$) is
\begin{equation}
  j_{\rm clust}(>N) = { L_* \Gamma[2+\alpha] \over \Gamma[1+\alpha,1] }
      \int_N^\infty  d N_{*666} N_{*666} { dn \over dN_{*666} }.
\end{equation}
The ratio $j_{\rm clust}/j_{\rm tot}$ is shown in Fig.~\ref{fig:lumfrac}
as a function of $N_{*666}$.
We find that the virialized regions of clusters more massive than
$10^{14}\,h^{-1}M_\odot$ contain $\simeq 7\%$ of the local luminosity.
This is quite comparable to the fraction of the mass contained in such
clusters in popular $\Lambda$CDM models with $\Omega_{\rm mat}\simeq 0.3$
and $\sigma_8\simeq 1$, though the number is particularly sensitive to
the latter assumption.  The enclosed fraction also depends on the density
contrast to which we extend the cluster edges.
For example, if we include galaxies out to $\Delta_N=100$, the luminosity
fraction in clusters doubles.

Since our data is consistent with $N(M)\sim M$ we would predict that
2MASS galaxies provide good tracers of the mass with a small and relatively
scale-independent bias on Mpc scales.
This would suggest that the correlation function (or power spectrum) of 2MASS
galaxies should show the virial `inflection' of the dark matter power spectrum
on Mpc scales, where the growth of fluctuations is first enhanced by non-linear
infall and then suppressed by virial motions within halos.  There is weak
evidence for this in Fig.~3 of Allgood, Blumenthal \& Primack (\cite{ABP}),
though further work with the full catalog needs to be done to really test this
prediction.
That $N(M)\sim M$ is perhaps not unexpected, as both simulations and
observations show that galaxies selected in the ``red'' trace the matter
distribution better than those selected in the ``blue''.
The original semi-analytic models (Kauffman et al.~\cite{KCDW}) predicted
a steeper slope for $N(M)$ for their red galaxies, with $N(M)\sim M^{0.9}$,
which is consistent with our results.

Theory predicts, and our simulations have assumed, that the number of galaxies
in a cluster mass halo should follow a Poisson distribution (at fixed mass).
It is difficult to test this with our current sample due to the difficulty
of estimating the mass of our clusters.
The ratio of $\langle N(N-1)\rangle^{1/2}/\langle N\rangle$, where $N$ is the
number of member galaxies determined from the matched filter probabilities
$p_i$, is very close to unity for clusters binned by $N_{*666}$.  (The Poisson
prediction is $\langle N(N-1)\rangle^{1/2}=\langle N\rangle$.)
However $N_{*666}$ is not a good proxy for $M_{200}$, having significant
scatter.  For the small number (110) of clusters for which we can detect all
$L_{*}$ galaxies and we have an X-ray temperature, the ratio
$\langle N(N-1)\rangle^{1/2}/\langle N\rangle$ is also consistent with unity,
though we need to choose broad bins in $T_X$ and the statistics are poor.

\section{Conclusions} \label{sec:conclusions}

We have applied the matched filter technique to both simulated galaxy
catalogs and the 2MASS galaxy catalogue to search for clusters over
approximately 90\% of the sky to a redshift limit of $z\simeq 0.05$.
We have matched our 2MASS derived catalog to existing catalogs in an
automated way using the NASA Extragalactic Database.
Our algorithm appears to be both robust and efficient, returning quite
complete samples of clusters out to distances where the typical cluster
contains as few as 3 galaxies.
The algorithm finds almost no `false groups', the main source of
contamination being the inclusion of objects below the mass cut in
the catalog.

The matched filter algorithm gives us a way of estimating cluster membership
and hence cluster properties like the multiplicity function, the number
function and the velocity dispersion in a new way.
While our estimates for the velocity dispersion agree well with earlier work,
we typically have far fewer galaxies per cluster than dedicated surveys
and hence larger errors.  This is offset by the fact that we have a large
number of clusters, over much of the sky, with which to search for correlations
between cluster properties such as velocity dispersion and X-ray temperature.
Where there is overlap we find quite good agreement with earlier work,
though often with improved statistics.

Although it is necessary to be very careful about the definition of $N$ in
order to compare with theory, we find that with sufficient care we have been
able to estimate the cluster number function, $dn/dN$, over more than 2 orders
of magnitude in $N$.

By using the velocity dispersion, X-ray luminosity or X-ray temperature as a
surrogate for mass we are able to estimate the multiplicity function $N(M)$.
Although there are serious issues in converting observables into cluster
masses, all of our results are fairly consistent and suggest that
$N(M)\sim M$ or perhaps slightly steeper, in reasonable agreement with earlier
estimates.
With all quantities normalized by the spherical radius corresponding to
a mass overdensity of $\Delta_M=200$ or the equivalent galaxy number
overdensity of $\Delta_N=200\Omega_M^{-1}=666$, we find that the number
of $L>L_*$ galaxies in a cluster of mass $M_{200}$ is 
\begin{equation}
  \log N_{*666} = (1.44\pm0.17)+(1.10\pm0.09)\log(M_{200}h/10^{15}M_\odot).
\end{equation} 
The uncertainties in this relation are largely due to the choice made
for relating the observed quantities to the cluster mass scale.  For
a fixed mass scale the scatter resulting from the different observed
correlations is considerably smaller.
Correlations of $N$ with other cluster properties, X-ray luminosity,
temperature or galaxy velocity dispersion, are given in
\S\ref{sec:multiplicity}.
The region has a K-band cluster mass-to-light ratio of $(M/L)_K=(116\pm46)h$
which is essentially independent of cluster mass.
The uncertainties are again dominated by the choice of the mass scale.
This scaling is consistent with $N(M)\sim M$, though if we take our best fit
seriously and $N(M)$ is steeper than $M$ we expect $M/L$ would fall slowly
with increasing mass.
Integrating over all clusters more massive than
$M_{200}=10^{14}\,h^{-1}M_\odot$, the virialized regions of clusters contain
7\% of the local stellar luminosity, quite comparable to the (somewhat theory
dependent) mass fraction in such objects in currently popular
$\Lambda$CDM models.

The cluster likelihoods tend to be larger in the real data than our mock
catalogs.
This difference could have several sources.
It could be due to the concentration of redshift measurements in the real
catalog towards groups and clusters rather than having the random distribution
of the synthetic catalog.  It may also indicate that the cosmology adopted in
the underlying simulation is incorrect (the number of clusters is very
sensitive to $\sigma_8$ for example), the normalization of the input $N(M)$
may be too low, our assumed $N(M)$ may have too many galaxies in low-mass
halos compared to high-mass halos, or the luminosity function could vary
systematically with the parent halo mass rather than being fixed
(as we have assumed).
Along these lines, both the expectation that light closely trace mass in
$K$-band and that the luminosity function change with mass suggest that
clusters contain a larger fraction of galaxies than we have assumed in our
current generation of mock catalogs.
These avenues will be explored in the future with larger simulations
and better data.

We view this work as but the first step in an iterative sequence.
Based on simulations which we know describe reality imperfectly, we have
calibrated our cluster finding algorithm.  When applied to the real data this
algorithm allows us to estimate correlations between different properties of
a halo which can be used in the next stage to improve the simulated catalogs.
As the 2MASS catalog becomes increasingly complete, and more redshifts become
available, we will estimate the global relations between different halo
parameters by providing them as priors to the fitting, and varying the
parameters to optimize the global likelihood.
These relations can then be used in the construction of the galaxy catalogs
{}from improved simulations which will allow us to further optimize and
understand the cluster finding algorithm.

\section*{Acknowledgments}
 
We would like to thank the 2MASS collaboration for their cooperation and
comments on this paper.
We would also like to thank N. Caldwell, M. Donahue, D. Eisenstein, C. Jones, 
M. Pahre, M. Postman and
K. Rines for helpful conversations and C. Baugh \& S. Cole for providing
additional IR properties for clusters in their semi-analytic models.
M.W. would like to thank H. Mo and R. Yan for enlightening
conversations on the halo model of galaxy clustering.
This publication makes use of data products from the Two Micron All Sky Survey,
which is a joint project of the University of Massachusetts and the
Infrared Processing and Analysis Center/California Institute of Technology,
funded by the National Aeronautics and Space Administration and the
National Science Foundation.
This research has made use of the NASA/IPAC Extragalactic Database (NED)
which is operated by the Jet Propulsion Laboratory, California Institute of
Technology, under contract with the National Aeronautics and Space
Administration.
C.S.K. and J.P.H. were supported by the Smithsonian Institution.
M.W. was supported by a Sloan Fellowship, the NSF and NASA.
Simulations were carried out at CPAC, through grants PHY-9507695,
and at the National Energy Research Scientific Computing Center.

\appendix

\section{Including Color Information}

In this appendix we outline the inclusion of color information in our 
algorithm.  To include color information in our algorithm we 
require a model for the multi-variate luminosity function in
luminosity and colors, $\phi(M,C_0)$, and its evolution with 
redshift.  The multi-variate and standard luminosity 
functions are related by $\phi(M)=\int \phi(M,C_0) dC_0$. Assuming
that the numbers of galaxies are not evolving, the measured
magnitudes and colors, $m$ and $c$ are related to the local 
values by $M=m-{\cal D}(z)$ (Eq.~\ref{eqn:absmag}) and
$C_0=c-{\cal C}(M,z)$ to remove the effects of distance,
K-corrections and evolution.

For the 2MASS survey the description simplifies further
because only the H--K color has a significant correlation 
with redshift and neither the H--K nor the J--H color
has a significant correlation with luminosity.  This
means that we can factor the bivariate luminosity 
function, $\phi(M,C_0)=\phi(M)\xi(C_0)$, with 
$\int \xi(C_0) dC_0 \equiv 1$.  The color distribution,
which consists of a narrow peak with a red tail, is well 
modeled by the sum of two Gaussians,
\begin{equation}
   \xi(H-K) =  
    {   f \over \sqrt{2\pi} \sigma_0 } e^{-(H-K-a_0)/2\sigma_0^2} +
    { 1-f \over \sqrt{2\pi} \sigma_1 } e^{-(H-K-a_1)/2\sigma_1^2}
\end{equation}
where $f=0.698$, $a_0=0.226$, $a_1=0.268$, $\sigma_0=0.025$
and $\sigma_1=0.081$.  The galaxies become steadily redder at 
higher redshift, with
\begin{equation}
   {\cal C}(z)=2.333 z - 0.770 z^2, 
\end{equation}
and it appears broader at fainter magnitudes due to measurement
errors that can be modeled by convolving the distribution with
a Gaussian of width $\sigma= 0.092 \times 10^{0.2(K-13)}$.  While the
errors in the flux are unimportant to our analysis given the
sample magnitude limit, the shallow slope of ${\cal C}(z)$ 
means that we must take into account the errors in the H--K
colors.

Given the bivariate luminosity function and the dependence
of the colors on redshift, we can modify the terms of the
likelihood function (see \S\ref{sec:fieldprob} and \S\ref{sec:clusterprob})
to include the additional information.  The probability of finding a field
galaxy with magnitude $m_i$, color $c_i$ and redshift $z_i$ is
\begin{equation}
   P_f(m_i,c_i,z_i) = 0.4\ln 10 D_C^2(z_i) {d D_c\over d z} 
      \phi_f(m-{\cal D}(z_i))\xi(c_i-{\cal C}(z_i)).
\end{equation}
Integrating over redshift, we obtain the number counts of galaxies
with a given color, which is equivalent to the probability of
finding a galaxy with magnitude $m_i$ and color $c_i$,
\begin{equation}
  P_f(m_i,c_i) = \int_0^\infty P_f(m_i,c_i,z) dz. 
\end{equation}
The probability distribution for an unknown galaxy redshift is
simply the ratio $P_f(m_i,c_i,z_i)/P_f(m_i,c_i)$.  
Because the slope of ${\cal C}$ is shallow, the addition of the
H--K color information only modestly improves the accuracy
of redshift estimates.  For the 50000 galaxies with known
redshifts, the observed errors in the predicted redshifts
are modestly smaller than their theoretical estimates of
$\sigma_z=0.028$ (using only fluxes), $0.036$ (using only colors) 
and $0.024$ (using both).  The errors in redshifts estimated using
colors (fluxes) grow slowly (linearly) with redshift, so at
low redshifts fluxes always become better redshift estimators 
than colors.  We must also modify the cluster membership
probabilities, replacing $\phi_c(M)$ by $\phi_c(M,C_0)$,
while leaving the normalization factor in the denominator,
$\Phi_c(M)$, unchanged.

This formalism for including color information in a matched filter algorithm
can be generalized to additional colors and more complicated descriptions for
the evolution of colors with redshift and luminosity, although tabulated
models will probably be required.  Matched filters can also exploit the redder
optical colors of cluster galaxies (as used in other algorithms,
e.g.~Goto et al.~\cite{Goto01}) by using a multi-variate cluster luminosity
function with a different color distribution from that for the field.

\end{document}